\documentclass[12pt]{article}

\pdfoutput=1

\usepackage{amssymb,amsfonts,amsthm}
\usepackage{graphicx}
\usepackage{subfigure}
\usepackage{amssymb}
\usepackage{amsthm}
\usepackage{amsmath}
\usepackage{bm}             
\usepackage[mathscr]{eucal}

\usepackage{cancel}

\usepackage{psfrag}

\usepackage{multirow}

\usepackage{wasysym}

\usepackage{color}

\def\hsp5{\hspace{5mm}}

\newcommand{\ts}{\mathsf{t}}
\newcommand{\xs}{\mathsf{x}}

\title{\sc On dynamical systems approaches and methods in $f(R)$ cosmology}

\begin{document}

\author{
\sc Artur Alho$^{1}$\thanks{Electronic address:{\tt aalho@math.ist.utl.pt}},\,
    Sante Carloni$^{2}$\thanks{Electronic address:{\tt sante.carloni@tecnico.ulisboa.pt}}\, and
    Claes Uggla$^{3}$\thanks{Electronic address:{\tt claes.uggla@kau.se}} \\
$^{1}${\small\em Center for Mathematical Analysis, Geometry and Dynamical Systems,}\\
{\small\em Instituto Superior T\'ecnico, Universidade de Lisboa,}\\
{\small\em Av. Rovisco Pais, 1049-001 Lisboa, Portugal.}\\
$^{2}${\small\em Centro Multidisciplinar de Astrofisica - CENTRA,}\\
{\small\em Instituto Superior T\'ecnico, Universidade de Lisboa,}\\
{\small\em Av. Rovisco Pais, 1049-001 Lisboa, Portugal.}\\
$^{3}${\small\em Department of Physics, Karlstad University,}\\
{\small\em S-65188 Karlstad, Sweden.}}
%
\date{}
\maketitle

\begin{abstract}

We discuss dynamical systems approaches and methods applied to flat Robertson-Walker models in
$f(R)$-gravity. We argue that a complete description of the solution space of a model requires
a global state space analysis that motivates globally covering state space adapted variables.
This is shown explicitly by an illustrative example, $f(R) = R + \alpha R^2$, $\alpha > 0$,
for which we introduce new regular dynamical systems on global compactly extended state spaces for
the Jordan and Einstein frames. This example also allows us to illustrate several local and
global dynamical systems techniques involving, e.g., blow ups of nilpotent fixed points, center
manifold analysis, averaging, and use of monotone functions. As a result of applying dynamical systems
methods to globally state space adapted dynamical systems formulations, we obtain pictures of the
entire solution spaces in both the Jordan and the Einstein frames. This shows, e.g., that due to
the domain of the conformal transformation between the Jordan and Einstein frames, not all the
solutions in the Jordan frame are completely contained in the Einstein frame. We also make comparisons
with previous dynamical systems approaches to $f(R)$ cosmology and discuss their advantages and
disadvantages.

\end{abstract}


\section{Introduction}

The simplest class of fourth order metric gravity theories is based on an action
\begin{equation}\label{actionJ}
\mathcal{S} = \int\left\{\frac{f(R)}{2\kappa^2} + \mathcal{L}_{m} \right\} \sqrt{-\det{g}}\, d^{4}x
\end{equation}
where $\kappa^2 = 8\pi G$; the speed of light, $c$, is set to one;
$\det{g}$ is the determinant of a Lorentzian 4-dimensional metric $g$,
and $R$ the associated curvature scalar, while $\mathcal{L}_{m}$ is the matter Lagrangian density.
General relativity with a cosmological constant $\Lambda$ is obtained by setting
$f(R)=R-2\Lambda$.

The vacuum part of these models, i.e., $\mathcal{L}_{m}=0$,
has recently achieved some popularity where
certain forms of the function $f(R)$ have resulted in geometric
models of inflation or, more recently, dark energy, see e.g.~\cite{capfra08}--\cite{aveetal16}
and also~\cite{sch07,barott83} for a historical background.
Although an assessment of cosmological viability requires a study of spatially
homogeneous and isotropic Robertson-Walker (RW) models and perturbations thereof,
we will restrict the analysis in this paper to flat RW cosmology. The
vacuum equations of these models can be written as
(see e.g.~\cite{LivRev10}):
\begin{subequations}\label{Jordanfulleq}
\begin{align}
\dot{a} &= Ha, \label{aeqf}\\
\dot{H} &= -2H^2 + \frac{R}{6}, \label{dotHf}\\
\ddot{R} &= -3H\dot{R} - \frac{1}{F_{,R}}\left[F_{,RR}\dot{R}^2 + \frac13\left(2f - FR\right) \right],  \label{ddotRf}
\end{align}
\begin{equation}
-6H\left(F_{,R}\dot{R} + FH\right) + FR - f = 0 .\label{constJorf}
\end{equation}
\end{subequations}
In the above equations an overdot represents the Jordan proper time $t$ derivative,
$a$ is the scale factor of the flat RW metric in the Jordan frame,
$H$ is the Jordan Hubble variable, $R$ is the Jordan curvature scalar, and
\begin{equation}
F = \frac{df}{dR}, \qquad F_{,R} = \frac{dF}{dR} = \frac{d^2f}{dR^2}, \qquad F_{,RR} = \frac{d^2F}{dR^2} = \frac{d^3f}{dR^3}.
\end{equation}
By regarding $\dot{R}$ as an independent variable, equation~\eqref{constJorf} forms a constraint
that the evolution equations must satisfy. Furthermore, equation~\eqref{aeqf} for $a$ decouples,
which leads to a reduced closed system of first order equations for $(H,\dot{R},R)$, which, due to the
constraint~\eqref{constJorf}, yield a dynamical system describing a flow on a 2-dimensional state space.
Once the reduced system of first order equations has been solved, the decoupled equation~\eqref{aeqf}
yields $a \propto \exp(\int dt H $).

The above system of equations has some general properties which are worth mentioning.
Firstly, the system is invariant under the transformation $(t,H) \rightarrow -(t,H)$, a property we will use below.
Secondly, the system is ill-defined whenever $F_{,R}=0$ for some value(s) of $R$.\footnote{Of course this is not
the case for general relativity for which $F_{,R}$ is identically zero, a case we will not consider here.}
This is related to pathological properties as regards the characteristics of $f(R)$ gravity, where, e.g., the
properties of gravitational waves will severely constrain the physical viability of such models.
It is therefore natural to divide $f(R)$ gravity into two main classes of models: those for
which $F_{,R}>0$ and those for which this is not the case. Thirdly, as it is well known $F>0$ is associated with that
one can introduce an Einstein frame. However, $F=0$ is not, in general, an invariant subset in the Jordan frame, since
\begin{equation}\label{dotFgen}
\dot{F}|_{F=0} = F_{,R}\dot{R}|_{F=0} = - \left. \frac{f}{6H}\right|_{F=0},
\end{equation}
where we have used~\eqref{constJorf} (i.e., $F=0$ is only an invariant subset if $f$ and $F$ are
simultaneously zero for some value of $R$). This suggest that it is also natural to divide $f(R)$ models
into two additional classes, those with $F>0$ everywhere, and those for which $F$ can change sign.
The latter case yields solutions in the Einstein frame that can be conformally extended in the
Jordan frame, but which ones depend on the explicit form of $f(R)$. 
We will later explicitly illustrate this result in the context of a specific model, which also exemplifies some
other general features of $f(R)$ cosmology.

There are a number of dynamical systems formulations in the literature that are based on transformations
from $(H,\dot{R},R)$ to some other variables (see Appendix~\ref{app:systems} for a discussion on several of these formulations).
In this context there has been considerable activity concerning fixed points (also called singular points, equilibrium points,
critical points) and their linear stability properties. It is therefore of interest to consider the
fixed points of the basic reduced state space variables $(H,\dot{R},R)$, which must satisfy
\begin{equation}\label{fixedgen}
-2H^2 + \frac{R}{6} = 0 ,\qquad \dot{R} = 0, \qquad 2f - FR =0,
\end{equation}
as follows from~\eqref{dotHf} and~\eqref{ddotRf}. Then,~\eqref{constJorf} takes the form
(assuming that $F_{,R}$ is non-zero)
\begin{equation}
-6H^2F + FR - f = 3F(-2H^2 + R/6)=0,
\end{equation}
which is thereby automatically satisfied. Moreover, $2f - FR$ is identically zero, if and only if $f \propto R^2$,
and only in this case there is a line of fixed points for which $R = 12H^2$, while all other models have isolated fixed points.
When transforming to other variables care has to be taken when it comes to the physical interpretation of fixed point results.
As we will see, some fixed points in other formulations simply reflect a break down of those variables, i.e., they
correspond to a state space coordinate singularity. It is also essential to note that fixed points do not always give
a complete asymptotic description. As will be emphasized in this paper, it is  necessary to consider the dynamics on
the entire state space of a given model to make an assessment of its physical content.

What is then required in order to obtain a complete description of the solution space and the properties of those
solutions for a given $f(R)$ RW model? We will illustrate some of the ingredients that are required to answer this
question with a specific example, but for all $f(R)$ RW cosmologies one needs to do the following:
\begin{itemize}
\item[(i)] State space analysis.
\item[(ii)] A complete state space adapted coordinate cover, including those state space boundaries for which
the equations can be extended (this e.g. excludes boundaries for which $F_{,R}$ becomes zero).
\item[(iii)] Local and global dynamical systems analysis.
\item[(iv)] Physical solution space interpretation.
\end{itemize}

Let us now comment on the above in a little more detail. (i) A state space analysis entails dimensional and scale considerations,
and a study of the algebraic structure of the constraint equation~\eqref{constJorf}, which includes global aspects
such as state space topology. (ii) This means that one needs to find
state space coordinates that \emph{globally} cover the state space of a given model, including the boundaries
for which the equations can be differentiable extended. This may include limits where $H$, $\dot{R}$ and $R$ become unbounded,
which motivates the introduction of new \emph{bounded} variables. Note that some models will even in
principle require several coordinate patches, but there are classes of models for which one can find common
\emph{local} useful variables. Even in cases where it is possible to find a bounded global state space coordinate system,
it might still be useful to consider other variables since it is unlikely that a global system, except under very special
circumstances, is the optimal one for all local structures, i.e., there might exist complementary sets of variables. Furthermore,
different models have different state space structures, and in general this requires different choices of variables --- the common
element is instead a state space analysis and an adaption to the structures that the analysis reveals.
(iii) To understand the solution space structure of a given model and the asymptotical behaviour of the solutions, which is essential for
assessing its physical viability (it is not enough to consider special solutions, e.g., fixed points), one must, in general, apply
linear \emph{and} non-linear fixed point techniques, as well as \emph{global} dynamical systems analysis. Furthermore, note
that fixed points will not in general give a complete asymptotic description, e.g., a problem might naturally give rise to 
limit cycles.
That a global understanding of the solution space is required is illustrated by the fact that if one has found a solution with a desirable
evolution, then the models will still only be of interest if this solution is in some sense an `attractor solution.' Even so,
this does not exclude that there exists an additional set of solutions that have a different evolution, which leads to issues
concerning measures describing how `typical' a solution is. 
(iv) Solutions, e.g. fixed points, have to be physically interpreted since a solution might be an artifact
of the variables one has used. For example, variables that do not cover the entire Jordan state space result in coordinate
singularities, which results in fixed points. Thus fixed points may not correspond to physical
phenomena, but may instead show that a formulation breaks down.

To illustrate the above issues (excluding the situation where $F_{,R}$ passes through zero, which we will comment on in
the final discussion), as well as allowing us to introduce some dynamical systems methods of quite wide applicability,
we will consider a specific example, the vacuum equations for the flat RW metric with
\begin{equation}\label{fR2}
f(R) = R + \alpha R^2, \qquad \alpha >0 .
\end{equation}
This model has attracted considerable attention in the past, see
e.g.,~\cite{Sta80}--\cite{gen15},
and it still remains as one of the more successful models of
inflation~\cite{plaXX15}. Although some interesting results have been obtained, previous analyses have been severely
hampered by formulations that do not give a complete, or sometimes correct, description of the global solution space
and its properties. In contrast, we will here give a complete description of the entire solution space of these models,
and we will also describe the solutions' asymptotic behaviour. More importantly though is that this model allows us
to explicitly address some aspects about how to obtain useful dynamical systems treatments of RW $f(R)$ models, and to
illustrate various dynamical systems methods. For example, we will situate the entire solution space of the Einstein
frame in the state space of the Jordan frame, which allows us to explicitly show how some solutions in the Einstein frame
for these models correspond to entire solutions in the Jordan frame, while other solutions can be conformally extended
in the Jordan frame. In other words, a local dictionary between the two frames does not always entail global equivalence
(the curious reader can skip ahead and take a look at Figures~\ref{fig:VacuumJF}, \ref{fig:Einsteinstatespace}
and~\ref{fig:Jordan_EinsteinBound} below).

The outline of the paper is as follows. In the next section we make a state space analysis
for the $f(R) = R + \alpha R^2$ models, which is used to produce a new regular
unconstrained dynamical system formulation on a compact state space for the Jordan frame.
We then use this system to perform a local analysis of the fixed points, focusing on
non-linear aspects such as blow ups of
nilpotent fixed points. This is followed by a global analysis that gives a complete
description of the entire solution space of the models, which is depicted and
summarized in Figure~\ref{fig:VacuumJF}. We emphasize the importance of the global topological
structure of the state space for a full understanding of the solution space.
In Section~\ref{sec:ein} we present a new regular unconstrained dynamical system
formulation on a compact state space for the Einstein frame. We then perform
a local analysis of fixed points, again focusing on non-linear aspects such as center
manifold analysis. It is also shown that the breakdown of the Einstein frame variables
at $F=0$ leads to fixed points in the Einstein frame state space that correspond to
coordinate singularities in the Jordan frame, thereby emphasizing the importance of physical interpretation
of fixed points. This is followed by global considerations, which yield a complete
description of the solution space in the Einstein frame. The section ends with
situating the global Einstein frame state space in the global Jordan frame
state space by means of the variable transformations that link the two approaches,
given in Appendix~\ref{app:relations}. This allows us to identify (a) the solutions in the
Einstein frame that can be conformally extended in the Jordan frame, and (b) the solutions
in the Einstein frame whose evolution completely describes that in the Jordan frame.
In Section~\ref{sec:disc} we comment on the relationship between our global Jordan state
space approach and other Jordan state space formulations, which are briefly reviewed in
Appendix~\ref{app:systems}, where their advantages and disadvantages are discussed. We also give
a fairly general discussion of $f(R)$ cosmology, which situates the present models in
this more general context.

\section{Dynamics in the Jordan frame}\label{sec:jor}

In this section we first perform a state space analysis of the $f(R) = R + \alpha R^2$, $\alpha >0$
vacuum models with flat RW geometry in the Jordan frame. The result is then used to derive a new
regular dynamical systems formulation on a global compactified state space, which, in contrast to
other formulations, completely covers the entire physical state space of these models, and its
asymptotic boundaries. We then use this state space picture to perform a local fixed
points analysis, which includes using blow up techniques and center manifold analysis, followed
by global considerations. This yields a complete description of the entire solution space,
depicted in Figure~\ref{fig:VacuumJF}.

\subsection{Dynamical systems formulation in the Jordan frame}\label{subsec:dynsysjord}

Specializing $f(R)$ to $f(R) = R + \alpha R^2$, $\alpha >0$, the evolution
equations~\eqref{dotHf} and~\eqref{ddotRf} can be written as:
\begin{subequations}\label{Jordanorigeq}
\begin{align}
\dot{H} &= -2H^2 + \frac{R}{6}, \label{dotH}\\
\ddot{R} &= -3H\dot{R} - \frac{R}{6\alpha},  \label{ddotR}
\end{align}
while the constraint~\eqref{constJorf} takes the form
\begin{equation}
-12H\left(\dot{R} + HR + \frac{H}{2\alpha}\right) + R^2 = 0 .\label{constJor}
\end{equation}
\end{subequations}

For future reference, note that restricting the general discussion leading to Eq.~\eqref{dotFgen},
which shows that the Einstein frame boundary $F=0$ is not in general an
invariant subset, to the present case,
\begin{equation}
F = 1 + 2\alpha R = 0 \qquad \Rightarrow \qquad R= - \frac{1}{2\alpha},
\end{equation}
yields that
\begin{equation}\label{Eframebound}
\dot{F}|_{F=0} = 2\alpha \dot{R}|_{F=0} = \frac{1}{24H\alpha},
\end{equation}
as also follows from~\eqref{constJor}. As a consequence there are solutions with $F>0$ that come from 
the region with $F<0$ (vice-versa if $H<0$)
and pass through the $F=0$ surface in the Jordan state space, i.e., some solutions in the
Einstein frame can be conformally extended in the Jordan frame (we will show this explicitly below).

Our first step in the state space analysis is to consider dimensions.
The dimensions of $t$, $H$, $R$, $\dot{R}$, and $\alpha$ are given by
$L$, $L^{-1}$, $L^{-2}$, $L^{-3}$, and $L^2$, respectively, where $L$ stands for length
(recall that the speed of light has been set to one). In contrast to general relativity,
the present models, which reflect a general feature of $f(R)$ gravity, break scale invariance.
As a consequence it will not be possible to use scale invariance to decouple an equation, as
is often done in dynamical systems treatments of general relativistic problems. However, we can
choose dimensionless variables that eliminate the explicit appearance of $\alpha$
(in general there can of course exist several dimensional parameters for which one can form
dimensionless ratios leaving a single dimensional parameter, where only the explicit appearance
of the latter can be eliminated by an appropriate choice of variables).

\begin{figure}[ht!]
\begin{center}
\includegraphics[width=0.5\textwidth,  trim = 0cm 2.0cm 0cm 0cm]{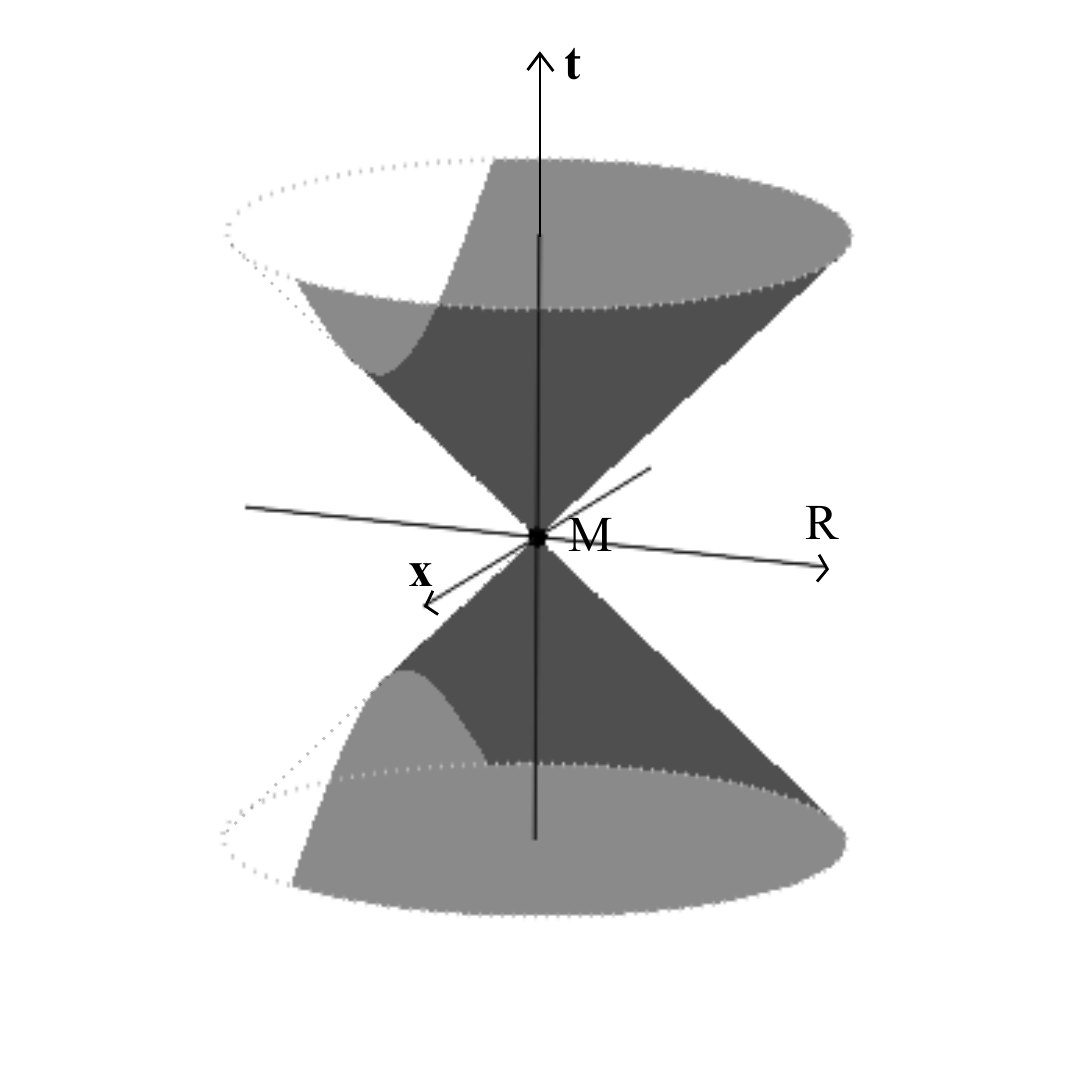}
\end{center}
\caption{The state space light cone for $f(R) = R + \alpha R^2$, $\alpha > 0$. The shaded part
denotes the state space domain of the Einstein frame, i.e., the state space of the Einstein frame
is a (non-invariant) subset of that of the Jordan frame.}
\label{fig:LightCone_SS}
\end{figure}

Our next step in our state space analysis is to study and simplify the constraint~\eqref{constJor}
as much as possible. For the present case it is possible to globally bring the constraint to a quadratic
canonical form where all variables have the same dimension. First note that if one chooses
$\dot{R} + HR + \frac{H}{2\alpha}$ as a new variable, then this variable as well as $H$ are seen
to be `state space null variables.' By appropriate scaling them with $\alpha$ so that they
obtain the same dimension $L^{-2}$ as $R$, and then making a linear transformation so that the constraint~\eqref{constJor}
takes a canonical quadratic form, results in
\begin{subequations}
\begin{align}
H &= \sqrt{\frac{\alpha}{12}}(\ts-\xs), \label{htx}\\
\dot{R}+ HR + \frac{H}{2\alpha} &= \frac{1}{\sqrt{12\alpha}}(\ts+\xs),
\end{align}
\end{subequations}
with
\begin{subequations}\label{tsxseq}
\begin{equation}\label{txconstr}
-\ts^2 + \xs^2 + R^2 = 0,
\end{equation}
where $\ts,\xs$ and $R$ all have dimension $L^{-2}$. It is important to note that the variable 
transformation $(H,\dot{R},R) \rightarrow (\ts,\xs,R)$ is globally valid since the Jacobian 
determinant is given by $1/6$.\footnote{The above state space structure is, of course, particular 
for the present models, but note that for models with $F_{,R}>0$ one can make a similar globally 
valid transformation which brings the constraint to the form $-\ts^2 + \xs^2 + g(R) = 0$, where 
$g(R)$ is determined by $f(R)$.} Thus the constraint
equation~\eqref{txconstr} makes it explicitly clear that the reduced vacuum state
space is a 2-dimensional double cone with a joint apex, see Figure~\ref{fig:LightCone_SS}.
The flow on this state space is determined by the following evolution equations:
\begin{align}\label{dimevol}
\dot{\ts} &= \frac{1}{2\sqrt{12\alpha}}\left(R - 2\alpha(\ts-\xs)^2\right), \\
\dot{\xs} &= \frac{1}{2\sqrt{12\alpha}}\left(-3R + 2\alpha(\ts-\xs)^2\right), \\
\dot{R} &= \frac{1}{2\sqrt{12\alpha}}\left(\ts + 3\xs - 2\alpha(\ts-\xs)R\right).
\end{align}
\end{subequations}
It follows from~\eqref{dimevol} that the two state space cones, defined by $\ts>0$ and $\ts<0$,
are disconnected invariant subsets with a fixed point $\ts=\xs=R=0$, as their common apex. This
fixed point, $\mathrm{M}$, represents the Minkowski solution, since $\ts=\xs=R=0 \Rightarrow H=\dot{R}=R=0$.
Note that $\mathrm{M}$ is the \emph{only} fixed point on the physical state space and
that it is non-hyperbolic\footnote{A non-hyperbolic fixed point is one for which a linearization yields eigenvalues
that not all have non-zero real parts.} (note that this is consistent with~\eqref{fixedgen} when specialized to the
present case).

Since the original system is invariant under the transformation
$(t,H) \rightarrow -(t,H)$, the system~\eqref{tsxseq}
is invariant under the transformation $(t,\ts,\xs) \rightarrow -(t,\ts,\xs)$.
It therefore suffices to investigate the dynamics on the invariant future
state space light cone with $\ts >0$ in order to obtain a complete picture of the dynamics.
Furthermore, the definition~\eqref{htx} in combination with the
constraint~\eqref{txconstr} implies that $H \geq 0$ on the future state space light
cone, i.e., it is arguably the future state space light cone that is of cosmological
interest. For these reasons we will only explicitly describe the dynamics on this
part of the global state space. It is clear from the above system that the minimum $H=0$ on the
future state space light cone $\ts>0$ only holds on the line $\ts = \xs$, $R=0$, but $H=R=0$
is not an invariant subset (except at $\mathrm{M}$). Indeed, since when $H=R=0$ and $\ts>0$
\begin{equation}
\dot{R} = \ts/\sqrt{3\alpha} >0,
\end{equation}
it follows that when $H=0$ then $R$ is passing through zero from negative to positive values.

To understand the present models it is essential to investigate if there are solutions that come from
the future state space null infinity. As a next step we therefore aim at producing a
regular system of equations on a compact state space. Furthermore, the variables need to be dimensionless so
that we eliminate the specific appearance of $\alpha>0$, which thereby automatically shows that this parameter
is not essential for the solution structure of the present models (this should not come as a surprise since 
it is possible to use units to set e.g. $\alpha=1$).


Removing the Minkowski fixed point $\ts=\xs=R=0$ from the analysis, we first introduce two new
dimensionless variables
\begin{equation}
(X,S) = \left(\frac{\xs}{\ts},-\frac{R}{\ts}\right),
\end{equation}
which are bounded thanks to the constraint~\eqref{txconstr}.
We then use that $\alpha\ts$ is dimensionless and positive on the future state space null cone
and introduce the bounded variable
\begin{equation}
T = \frac{1}{1 + 2\alpha\ts}.
\end{equation}
Next, to obtain a regular dimensionless system of evolution equations we introduce
a new dimensionless time variable $\bar{t}$, defined by
\begin{equation}
\frac{dt}{d\bar{t}} = 2\sqrt{12\alpha}T,
\end{equation}
which leads to
\begin{subequations}\label{dynsysTXS}
\begin{align}
T^{\prime} & = T(1-T)\left[TS + (1-T)(1-X)^2\right],\\
X^{\prime} &= S\left[T(3+X) + (1-T)(1-X)S\right], \\
S^{\prime} &= -X\left[T(3+X)+ (1-T)(1-X)S\right],
\end{align}
where ${}^\prime$ denotes the derivative with respect to $\bar{t}$,
subjected to the constraint
\begin{equation}\label{XSconstr}
X^2 + S^2 = 1 .
\end{equation}
\end{subequations}

Note that the above variable change from $(\ts, \xs, R)$ to $(T, X, S)$
amounts to a projection where all circles on the light cone with constant
$\ts$ now become the unit circle given by $X^2 + S^2 =1$, where the different circles
are parameterized by the value of $T$, i.e, the above variables cover all of the future
state space light cone.\footnote{The sign in the definition of $S$ has been chosen in order to simplify 
the comparison with the Einstein frame state space. The reason for
defining $T$ as a monotonically decreasing function of $\ts$ instead of a
monotonically increasing one, e.g., by setting $T=2\alpha \ts/(1 + 2\alpha\ts)$, is that
the Minkowski state, as we will prove, is the future asymptotic state of all solutions,
and that this definition also makes the transition to the variables we use to describe the
Einstein frame state space more convenient.} The present state space ${\bf S}$, which is just the
future state space light cone, is given by a finite cylinder determined by
\begin{equation}
0 < T <1, \qquad  X^2 + S^2 = 1.
\end{equation}

Because the state space ${\bf S}$ is relatively compact (i.e. its closure is compact) and
the equations are completely regular, we can extend the state space ${\bf S}$ to include the
invariant boundaries $T=0$, and $T=1$
to obtain an extended compact state space $\bar{\bf S}$. This turns out to be essential since, as
we will see, the asymptotic states for all solutions within the physical state space ${\bf S}$ reside
on these invariant boundary subsets, see Figure~\ref{fig:VacuumJF} below.
Indeed, we will prove that there are no fixed points or periodic orbits
in the physical interior state space. Thus, all solutions in ${\bf S}$ originate
from fixed points on $T=0$ (the future null infinity of the future state space light cone),
and end at a limit cycle on $T=1$ (which describes how all solutions asymptotically
approach the Minkowski space-time). Thus the present variables represent a compactification
of the future state space light cone, where they blow up the neighborhood of the non-hyperbolic
Minkowski fixed point in the dynamical system for $\ts,\xs, R$, thereby yielding a correct description of how
all solutions approach the future asymptotic Minkowski state. 

It is of interest to express $(H,\dot{R},R)$ in terms of $(T,X,S)$:
\begin{subequations}
\begin{align}
H &= \frac{1}{2\sqrt{12\alpha}}\frac{(1-T)(1-X)}{T},\label{HJ}\\
\dot{R} &= \frac{1}{4\alpha\sqrt{12\alpha}}\frac{(1-T)}{T}\left(1 + 3X + \frac{(1-T)(1-X)S}{T}\right),\\
R &= -\frac{1}{2\alpha}\frac{(1-T)S}{T}.
\end{align}
\end{subequations}
%
%
These equations reveal that $H$ is not only zero on the invariant boundary $T=1$, but also when $X=1$, $S=0$, which
corresponds to the line $\ts = \xs$, $R=0$ on the future state space light cone. Since this is not an
invariant subset of the dynamical system~\eqref{dynsysTXS}, the solution trajectories pass through $X=1$,
going from positive to negative $S$. Indeed, we will show later on that all solutions pass through $X=1$, $S=0$
infinitely many times.

Finally, although~\eqref{dynsysTXS} is a constrained system, the constraint~\eqref{XSconstr} is easily globally solved
by introducing
\begin{equation}
X = \cos{\theta},\qquad S = \sin{\theta},
\end{equation}
which results in the following unconstrained regular system of equations:
\begin{subequations}\label{Jordan2DDynSys}
\begin{align}
T^{\prime} &= T(1 - T)\left[T\sin{\theta} + (1-T)(1 - \cos{\theta})^2\right], \\
\theta^{\prime} &= -T(3 + \cos{\theta}) - (1-T)(1 - \cos{\theta})\sin{\theta}.
\end{align}
\end{subequations}
The above regular global dynamical system  form our `master equations' for dealing with the
present models in the Jordan frame. However, since the present formulation differs substantially
from the ones in the literature it is of interest to take a look at some other formulations
for the Jordan frame and make comparisons, which we do in Section~\ref{sec:disc} and in
Appendix~\ref{app:systems}. Finally, we stress that the above system was possible because we
adapted the variables to the particular state space properties of the present models; other
models need different variables. However, to find one (or more) set(s) of (differentiably overlapping) 
useful variables covering the entire state space and its possible infinite limits, one needs to go 
through the same steps of (i) state space analysis and (ii) state space adapted coordinates as for 
the present illustrative example. Next we turn to illustrating (iii): local and global dynamical systems analysis.




\subsection{Local fixed point analysis in the Jordan frame}\label{subsec:localjord}


In this subsection we perform a local analysis of the fixed points of our
new regular dynamical system~\eqref{Jordan2DDynSys} on the compactified
global state space, with a focus on necessary non-linear aspects.
As we will prove below, all fixed points are located
on the boundary subset $T=0$, associated with $H \rightarrow \infty$.
Considering this subset, we find that there are two fixed points:
\begin{subequations}\label{FPvacumJordan}
\begin{align}
\mathrm{R}\!: \quad \theta &= \pi + 2n\pi, \\
\bar{\mathrm{dS}}\!:\quad \theta &= 2n\pi,
\end{align}
\end{subequations}
with $n$ an integer. The motivation for the nomenclature for these fixed points will be made clear below.

The fixed point $\mathrm{R}$ is a hyperbolic source, while $\bar{\mathrm{dS}}$ has two zero
eigenvalues. More precisely, it is nilpotent of first degree. Such fixed points are dealt with
by means of so-called blow up techniques, described in detail in~\cite{andetal71} and~\cite{dumetal06}.
In order to bring the problem to standard form for nilpotent fixed points we
first scale the variables $\theta$ and $T$ and introduce the following notation (without loss
of generality, we choose the representation $\theta = 0$ for the fixed point $\bar{\mathrm{dS}}$):
\begin{equation}
x = -\theta, \qquad y = 4T.
\end{equation}
This leads to a dynamical system on the form
\begin{subequations}
\begin{xalignat}{2}
x^{\prime} &= y + P(x,y); &\quad P(x,y) &= a(x) + b(x)y,\\
y^{\prime} &= Q(x,y); &\quad Q(x,y) &= c(x)y + d(x)y^2 + e(x)y^3,
\end{xalignat}
\end{subequations}
where
\begin{subequations}
\begin{xalignat}{2}
a(x) &= -(1-\cos{x})\sin{x}, &\quad b(x) &= -\frac{1}{4}(1-\cos{x})\left(1-\sin{x}\right), \\
c(x) &= (1-\cos{x})^2, &\quad d(x) &= -\frac{1}{4}\left(\sin(x)+2(1-\cos{x})^2\right),  \\
e(x)&= \frac{1}{16}\left(\sin{x}+(1-\cos{x})^2\right). & \quad &
\end{xalignat}
\end{subequations}
Next we introduce a new variable $Y$ instead of $y$:
\begin{equation}
Y = y + P(x,y)= a(x) + (1+b(x))y,
\end{equation}
which leads to
\begin{equation}\label{yY}
y = \frac{Y + (1-\cos{x})\sin{x}}{1-\frac{1}{4}(1-\cos{x})\left(1-\sin{x}\right)}.
\end{equation}
In the neighborhood of the origin this means that the dynamical system takes the form
\begin{subequations}
\begin{align}
x^{\prime} &= Y, \\
Y^{\prime} &= \frac{x^7}{16}\left(1+h(x)\right) - \frac{3x^2}{2}\left(1+g(x)\right)Y + j(x,Y)Y^2,
\end{align}
\end{subequations}
where
\begin{subequations}
\begin{align}
h(x) &= -\frac{7}{24}x^2 - \frac{1}{4}x^3 + \dots, \\
g(x) &= -\frac{1}{3}x^2 - \frac{1}{8}x^3 + \dots, \\
j(x,Y) &= -\frac{1}{2}x + \frac{3}{8}x^2+ \dots + \left(\frac{1}{16}x + \dots\right)Y.
\end{align}
\end{subequations}
We now proceed by making the following so-called blow-up transformation
\begin{equation}
(x,Y) = (u,u^3 \bar{y}),
\end{equation}
and change time variable by dividing the right hand sides by $u^2$. This results in
\begin{subequations}
\begin{align}
u^{\prime} &= \bar{y}u, \\
\bar{y}^{\prime} &= -\frac{3}{2}\bar{y}\left(1 + 2\bar{y} + f(u,\bar{y})\right)
+ \frac{u^2}{16}\left(1+h(u)\right),
\end{align}
\end{subequations}
where the ${}^\prime$ now refers to the new time variable and where
\begin{equation}
f(u,\bar{y}) = g(u) - \frac{2u}{3}j(u,\bar{y})\bar{y},
\end{equation}
which obeys $f(0,\bar{y})=0$ and
$\frac{\partial f}{\partial u}(0,\bar{y})=\frac{\partial f}{\partial \bar{y}}(u,0) = 0$.

It follows that on the $u=0$ subset there are two fixed points
\begin{subequations}
\begin{alignat}{2}
\mathrm{S}\!: \quad \bar{y} &= -\frac{1}{2}, \\
\mathrm{dS}\!:\quad \bar{y} &= 0.
\end{alignat}
\end{subequations}
The fixed point $\mathrm{S}$ is a hyperbolic saddle 
while $\mathrm{dS}$ is a non-hyperbolic fixed point
with eigenvalues zero and $-3/2$.

To deal with $\mathrm{dS}$ we apply center manifold theory (for examples of center manifold
analysis in cosmology, see e.g.~\cite{alhugg15a}--\cite{tam14}).
The center manifold $W^c$ can be obtained as the graph $\bar{y}=\varphi(u)$ near $(u,\bar{y}) = (0,0)$
(i.e., use $u$ as an independent variable), where $\varphi(0) = 0$ (fixed point
condition) and $\frac{d\varphi}{du}(0) = 0$ (tangency condition).
This leads to
\begin{equation}
-\frac{3}{2}\varphi(u)\left[1 + 2\varphi(u)+ f(u,\varphi(u))\right]
+ \frac{u^2}{16}(1 + h(u)) - u\varphi(u)\frac{d\varphi}{du} = 0.
\end{equation}
This differential equation can be solved approximately by representing $\varphi(u)$ as the formal power series
\begin{equation}
\varphi(u) = \sum_{i=2}^n a_i u^i + {\cal O}(u^{n+1}) \qquad \text{as}\qquad u\rightarrow 0.
\end{equation}

\begin{figure}[ht!]
\begin{center}
\includegraphics[width=0.5\textwidth]{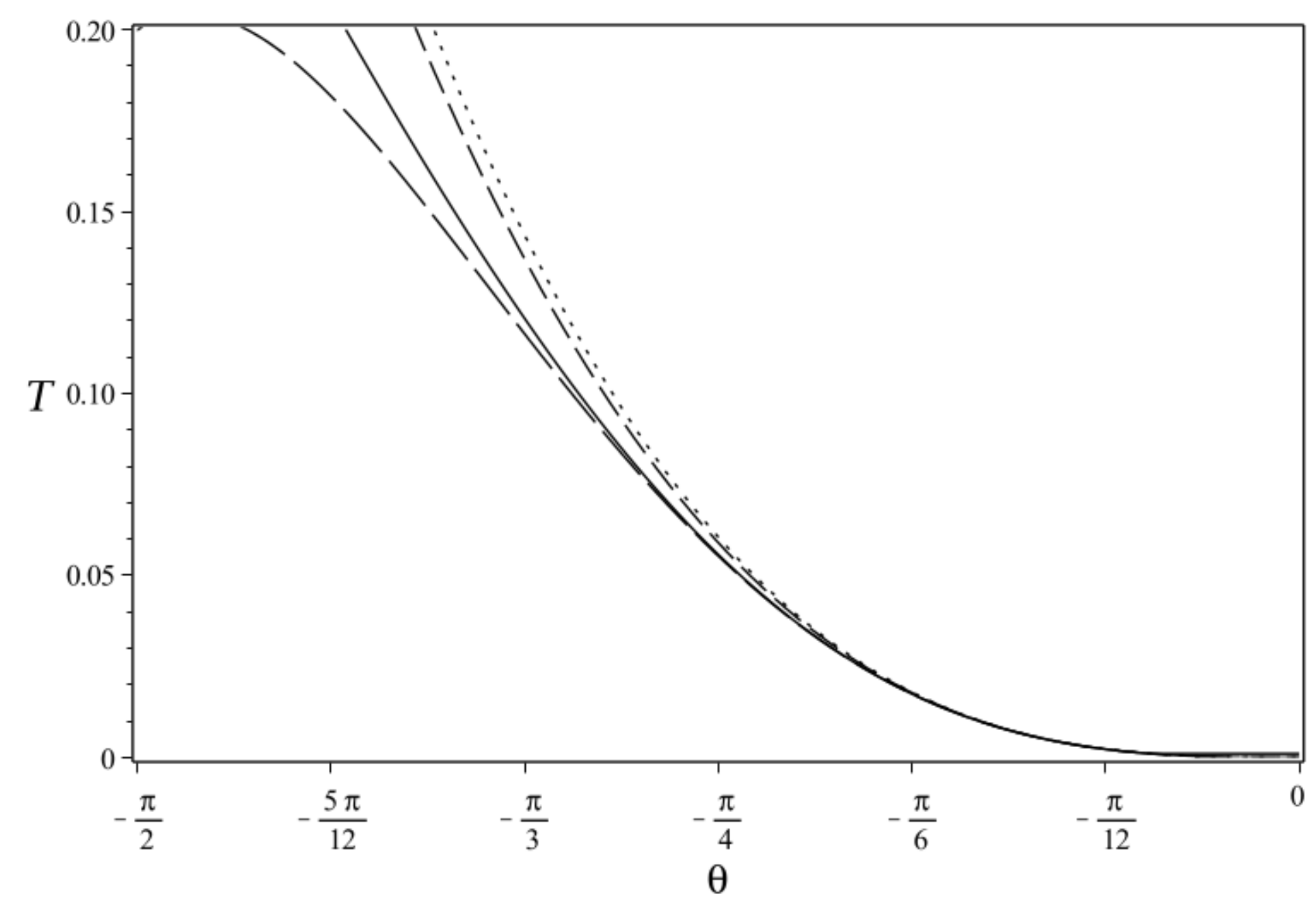}
\end{center}
\caption{Comparisons of the center manifold expansion of $\mathrm{dS}$ with the
numerically computed solution given by the solid line. The
leading-order term in the center manifold expansion is given by the dotted line; the
leading-order correction to this by the dashed line, and the next order correction by the long-dashed line.}
\label{fig:JordanCM}
\end{figure}

Solving algebraically for the coefficients we find
%
\begin{equation}
\varphi(u) = \frac{u^2}{24}\left(1 - \frac{7}{72}u^2 + \dots\right) \qquad \Rightarrow \qquad
Y = \frac{u^{5}}{24}\left(1-\frac{7}{72}u^2 + \dots\right)\, .
\end{equation}
The present case corresponds to Figure 3.16 (a) on p. 112 in~\cite{dumetal06}. The saddle
$\mathrm{S}$ is associated with orbits (i.e., solution trajectories) that approach
$\bar{\mathrm{dS}}$ from the region $T\leq0$ while the center manifold of
$\mathrm{dS}$ with $T>0$ corresponds to the only solution from $\bar{\mathrm{dS}}$ that
enters the physical state space. 
Inserting the above expression for $Y$ into~\eqref{yY} leads to the expression (note that $\theta <0$)
\begin{equation}\label{thetaexodS}
T = -\left(\frac{\theta}{2}\right)^3\left[1 - \frac16\left(\frac{\theta}{2}\right)^2 + \left(\frac{\theta}{2}\right)^3 + \dots\right].
\end{equation}
This is a series expansion that approximates the `inflationary attractor solution' that enters the physical state space
from $\bar{\mathrm{dS}}$.
The accuracy of this approximation compared to the numerical solution can be found in Figure~\ref{fig:JordanCM}.
If one is so inclined, one can obtain further approximation
improvements by means of so-called Pad{\'e} approximants, as described in e.g.~\cite{alhugg15a,alhugg15}, and
references therein.

Finally, note that the deceleration parameter $q$ in the Jordan frame,
defined by $dH/dt=-(1+q)H^2$, is given by
\begin{equation}\label{qJ}
q = 1 + 4\left(\frac{T}{1-T}\right)\frac{\sin\theta}{(1 - \cos\theta)^2}.
\end{equation}
It follows that, except at $\bar{\mathrm{dS}}$, where $\cos\theta=1$ and~\eqref{qJ} is ill-defined,
the deceleration parameter takes the value $q=1$ on the invariant boundary $T=0$, including the fixed point $\mathrm{R}$.
The reason for choosing this notation for the fixed point is due to the fact that this value of $q$ corresponds to a Universe filled with
radiation in general relativity.
On the other hand, the fixed point $\bar{\mathrm{dS}}$ does not describe the asymptotic features of the solution
that originates from it into the physical state space, since the right hand side of~\eqref{qJ} diverges for $\theta=2n\pi$.
However, inserting the asymptotic expression~\eqref{thetaexodS} for the center manifold of $\mathrm{dS}$ into~\eqref{qJ} leads
to the following expansion in $\theta$:
\begin{equation}
q = -1 + \frac{1}{12}\theta^2 + \frac{1}{120}\theta^4 + \dots
\end{equation}
which reveals that the center manifold solution has $q=-1$ asymptotically, i.e., `the inflationary attractor solution'
originates from a (quasi) de-Sitter state. Furthermore,
just as for the solutions originating from $\mathrm{R}$, this state is associated with
$H\rightarrow \infty$, as follows from~\eqref{HJ} and~\eqref{thetaexodS}.

In the next section we will prove that the one-parameter set of solutions that enter the
physical state space from $\mathrm{R}$, and the single solution that comes from $\bar{\mathrm{dS}}$,
all of them originating from a singular state at $H \rightarrow \infty$ (i.e., at future null
infinity with respect to the light cone state space), constitute all solutions in the physical state space.



\subsection{Global analysis in the Jordan frame}\label{subsec:globaljord2}

Consider the function
\begin{equation}
J = \frac{(1-T)(3+\cos{\theta})}{T} > 0, \label{DefJ}
\end{equation}
which obeys the equation
\begin{equation}
J^{\prime} = -2\frac{(1-T)^2(1-\cos{\theta})^2}{T} .
\end{equation}
It follows that $J$ is monotonically decreasing when $0<T<1$ and $\theta\neq 2n\pi$.
Furthermore, since
\begin{equation}
J^{\prime\prime}|_{\theta=2n\pi} = 0, \qquad J^{\prime\prime\prime}|_{\theta=2n\pi} = 0,\qquad
J^{\prime\prime\prime\prime}|_{\theta=2n\pi} = -4^4 (1-T)^2T,
\end{equation}
it follows that $\cos\theta=1$ only represents an inflection point in the evolution
of $J$. As a consequence $J \rightarrow \infty$ when $\bar{t} \rightarrow - \infty$,
which implies that all orbits in the physical state space ${\bf S}$ with $0<T<1$
originate from the subset $T=0$, while $J \rightarrow0$ when
$\bar{t} \rightarrow + \infty$, which implies that all
orbits in the physical state space $0<T<1$ end at the subset $T=1$. There are thereby no fixed
points or periodic orbits in the physical state space ${\bf S}$.

The analysis of the subset $T=0$ is trivial and our previous investigation of the fixed
points on $T=0$ shows that there is a single orbit that enters the physical state space from
$\bar{\mathrm{dS}}$ while there is a 1-parameter set that originates from $\mathrm{R}$.
The above global considerations based on $J$ proves that these local fixed point results
describe the origins of \emph{all} solutions in the physical state space ${\bf S}$.

The invariant subset $T=1$ yields the equation
\begin{equation}
\theta^{\prime} = -(3 + \cos{\theta}),
\end{equation}
as follows from~\eqref{Jordan2DDynSys}. Since $3 + \cos{\theta}>0$ it follows that
$T=1$ represents a periodic orbit where $\theta$ is monotonically
decreasing. From our considerations of the function $J$, this proves that this periodic
orbit is a limit cycle that describes the future asymptotic behaviour of \emph{all} solutions
in the physical state space ${\bf S}$, i.e., it constitutes the $\omega$-limit set of all
solutions with $0<T<1$. As an aside, this provides a simple cosmological example that it
is often not sufficient to just do fixed point analysis.

\begin{figure}[ht!]
\centering
\subfigure[Solutions on the Jordan frame state space cylinder.]{\label{fig:jorcyl}
\includegraphics[width=0.45\textwidth, trim = 0cm 0.25cm 0cm 0cm]{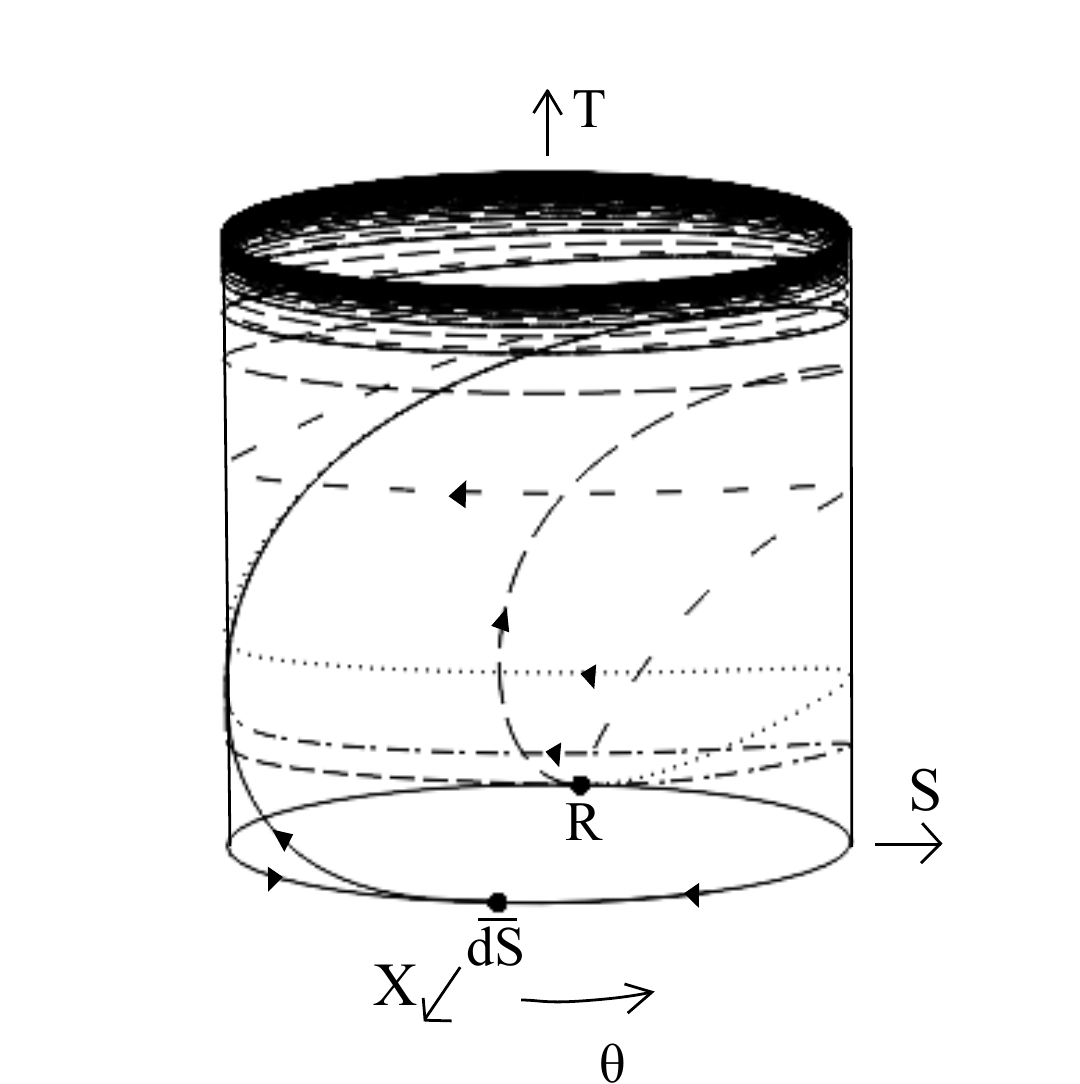}}
\subfigure[Solutions on the unwrapped Jordan state space cylinder.]{\label{fig:jorunwrapped}
\includegraphics[width=0.40\textwidth]{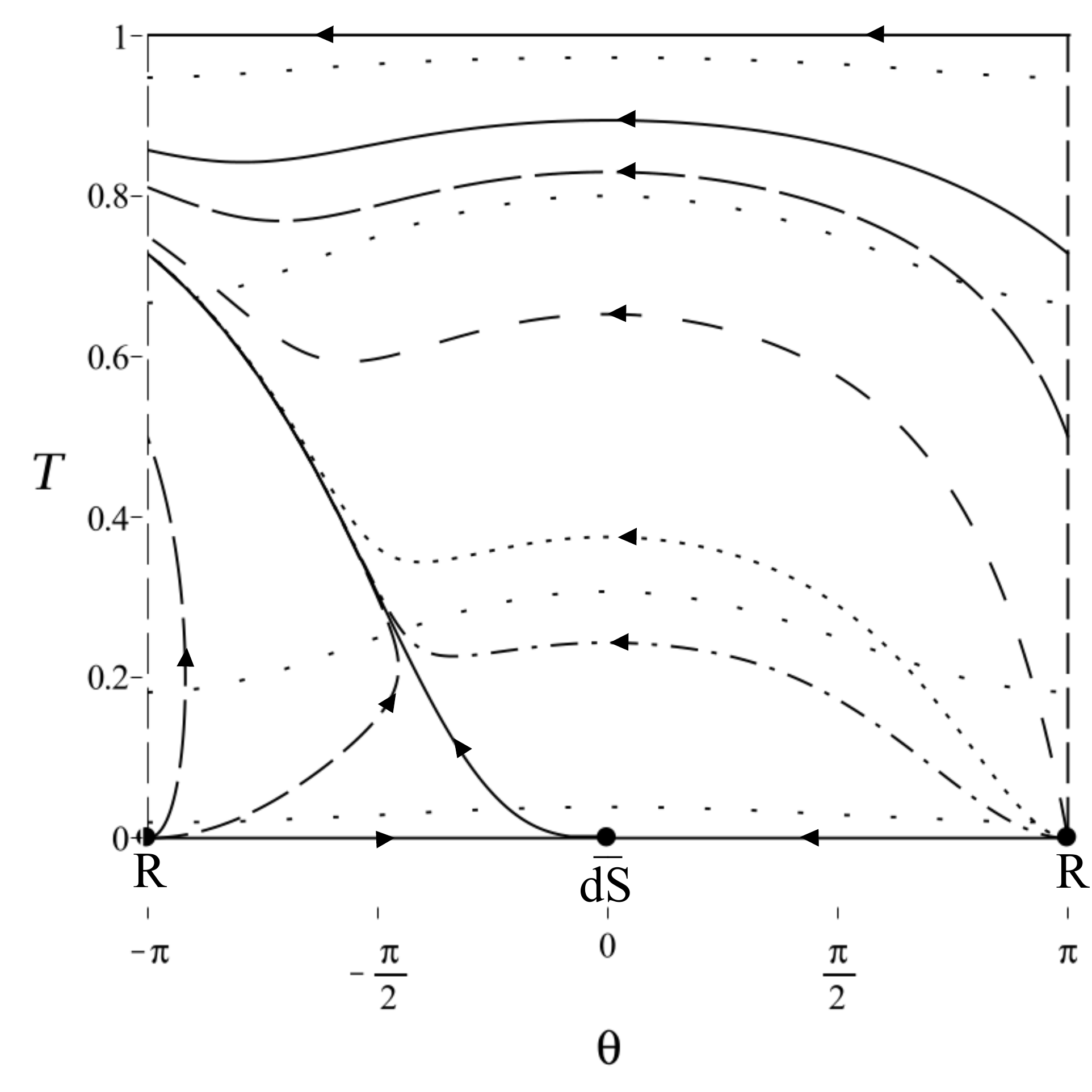}}
\caption{Two representations of the Jordan frame state space. All solutions in the Jordan frame state space end
at the periodic orbit at $T=1$, and they all originate from the fixed point
$\mathrm{R}$, except for `the inflationary attractor solution' (solid line) that comes from $\bar{\mathrm{dS}}$.
The space-dotted lines in Figure (b) depict constant values of the monotone function $J$.}
\label{fig:VacuumJF}
\end{figure}

We end this section by depicting representative solutions describing the entire solution space in the
Jordan frame in Figure~\ref{fig:VacuumJF}. Note that there is an open set
of solutions that are not attracted to the inflationary attractor solution until the
oscillatory regime at late times, where all solutions approach the future attractor,
i.e., the limit cycle at $T=1$. Thus, to argue that the inflationary attractor solution is in some sense
an attractor requires the introduction of some measure. In this context
we refer to the recent interesting discussion about scales and measures given in~\cite{slo16}, and references
therein.

As a final remark, we note that heuristic approximations in the Jordan frame for the inflationary attractor solution
and for the oscillatory `reheating' regime at late times have been given in~\cite{mijetal86}, and later reproduced in~\cite{LivRev10}.
Next we will deal with  Einstein frame dynamics, and then we will present rigorous approximations schemes for the oscillatory regime at late times.
Such methods can also be applied to the Jordan frame,
or one can translate the approximations in the Einstein frame to the Jordan frame by means of the relations given in
Appendix~\ref{app:relations}, but for brevity we will refrain from doing this.

\section{Dynamics in the Einstein frame}\label{sec:ein}

The analysis in the Einstein frame will serve as an illustrative example of (iv): physical solution space interpretation.
It exemplifies the situation where a state space only covers part of the Jordan frame state space,
which, e.g., leads to coordinate singularities in the form of fixed points. Thus fixed points in a
given formulation may not correspond to physical phenomena, but may instead reflect that the
formulation breaks down, thus necessitating a physical interpretation. (We will see several other
examples of this in
appendix~\ref{app:systems}.) When dealing with the Einstein frame we will introduce a new regular
unconstrained dynamical system on a compact state space in the Einstein frame, which gives a complete
description of the solution space in this frame. This will enable us to situate the entire solution
space in the Einstein frame in the state space of the Jordan frame.

\subsection{Dynamical systems formulation in the Einstein frame}\label{subsec:dynsysein}

The Einstein frame formulation of $f(R)$ gravity is based on the following
conformal transformation of the Jordan metric $g_{\mu\nu}$ to the Einstein frame metric
$\tilde{g}_{\mu\nu}$ (see e.g.~\cite{LivRev10} and~\cite{barcot88}):
\begin{equation}
\tilde{g}_{\mu\nu} = F g_{\mu\nu}, \qquad F = \frac{df}{dR},
\end{equation}
which thereby assumes that $F>0$. Thus $F=0$ constitutes the boundary between the Einstein and the Jordan frame state spaces,
of which it in general will be a subset,
but not an invariant subset in the Jordan frame, as shown in eq.~\eqref{Eframebound}. As a consequence, as we
will see, there are solutions with $F>0$ that come from the region with $F<0$ and
pass through the $F=0$ surface in the Jordan state space, i.e., some solutions
in the Einstein frame can be (conformally) extended in the Jordan frame.

Introducing
\begin{equation}\label{EFscalarfield}
\kappa\phi = \sqrt{\frac{3}{2}}\ln F, \qquad V(\phi) = \frac{RF - f}{2\kappa^2F^2}
\end{equation}
the action in the Jordan frame~\eqref{actionJ} transforms to an action with the Einstein-Hilbert form
for a scalar field minimally coupled to gravity,
%
\begin{equation}\label{EHaction}
\mathcal{S} = \int \left\{\frac{\tilde{R}}{2\kappa^2}-\frac{\tilde{g}^{\mu\nu}}{2}\left(\nabla_{\mu}\phi\right)
\left(\nabla_{\nu}\phi\right)- V(\phi)+ F^{-2}(\phi)\tilde{\mathcal{L}}_{m} \right\} \sqrt{-\det{\tilde{g}}}\, d^{4}x,
\end{equation}
where $\tilde{R}$ is the curvature scalar of the Einstein frame metric $\tilde{g}_{\mu\nu}$.

\begin{figure}[ht!]
\begin{center}
\includegraphics[width=0.6\textwidth]{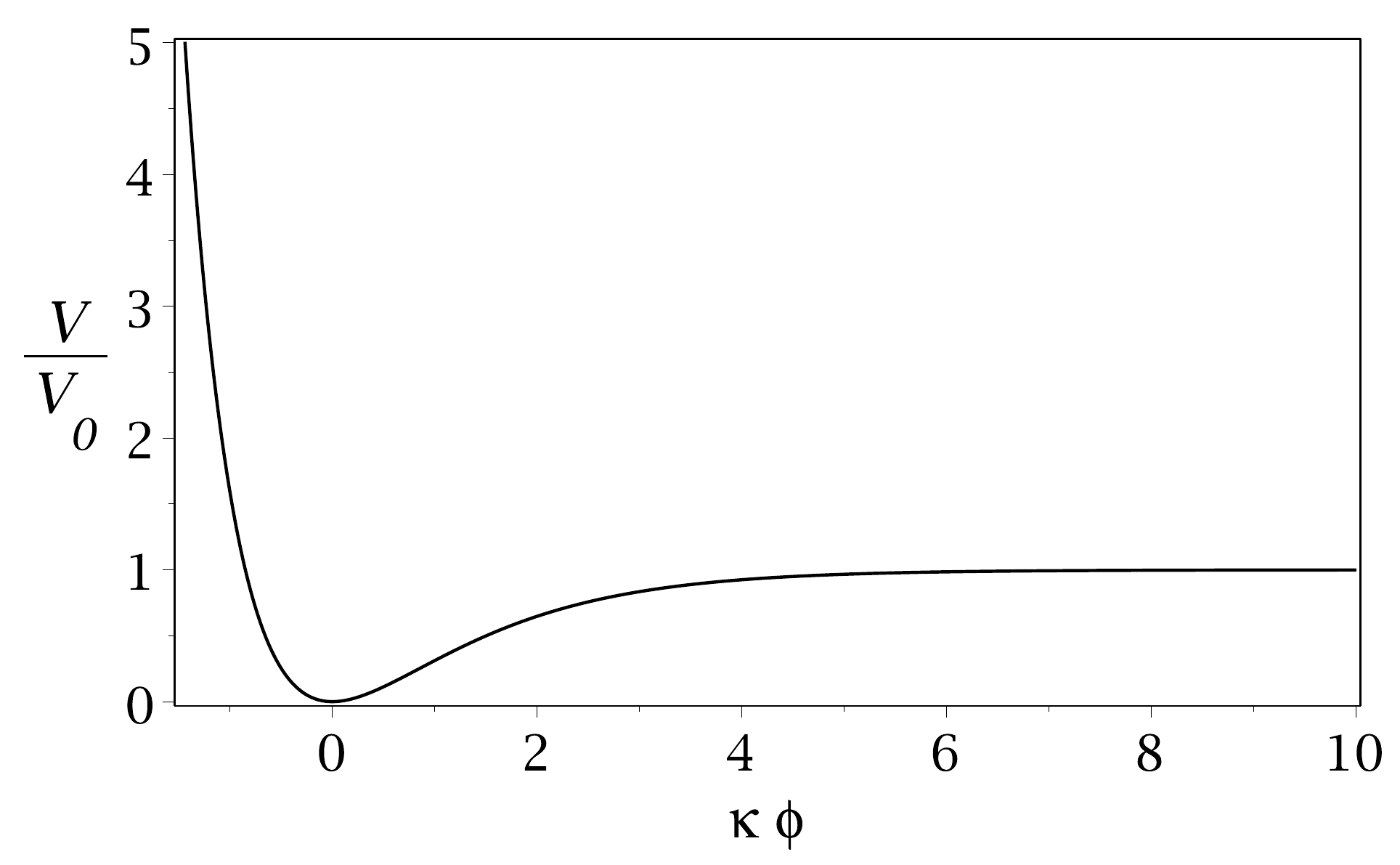}
\end{center}
\caption{The potential $V(\phi)$ for the minimally coupled scalar field in the Einstein frame corresponding to the model
$f(R)=R+\alpha R^2$, $\alpha > 0$.}
\label{fig:Potential}
\end{figure}

Specializing to the present $f(R)=R+\alpha R^2$ vacuum models leads to (see e.g.~\cite{bar88})
\begin{equation}
\kappa\phi = \sqrt{\frac{3}{2}}\ln(1 + 2\alpha R), \qquad V(\phi) = V_{0}\left(1-e^{-\sqrt{\frac{2}{3}}\kappa\phi}\right)^{2}\,,
\end{equation}
where
\begin{equation}
V_0=\frac{1}{8\alpha\kappa^2}>0 .
\end{equation}
The above potential is depicted in Figure~\ref{fig:Potential}.

In a flat RW geometry, the present models in the Einstein frame yield the following evolution equations
\begin{subequations}
\begin{align}
\frac{d\tilde{a}}{d\tilde{t}} &= \tilde{H}\tilde{a}, \label{Edota}\\
\frac{d\tilde{H}}{d\tilde{t}} & = - \frac{\kappa^2}{2}
\left(\frac{d\phi}{d\tilde{t}}\right)^2, \label{Ray1} \\
\frac{d^2\phi}{d\tilde{t}^2} &= -3\tilde{H}\frac{d\phi}{d\tilde{t}}
- 2\sqrt{\frac{2}{3}}\kappa V_{0}\left(1 - e^{-\sqrt{\frac{2}{3}}\kappa\phi}\right) e^{-\sqrt{\frac{2}{3}}\kappa\phi}, \label{waveEq}
\end{align}
and the constraint
\begin{equation} \label{Gauss1}
3\tilde{H}^2 =\kappa^2
\left[\frac12 \left(\frac{d\phi}{d\tilde{t}}\right)^2 + V_{0}\left(1-e^{-\sqrt{\frac{2}{3}}\kappa\phi}\right)^{2}\right],
\end{equation}
\end{subequations}
where $\tilde{t}$ denotes the Einstein frame proper time variable, while
$\tilde{a}$ and $\tilde{H}$ are the Einstein frame scale factor and Hubble variable,
respectively.

As in the Jordan frame, the Einstein frame scale factor $\tilde{a}$ decouples,
leaving a reduced closed system of first order evolution equations for
$(\tilde{H},\frac{d\phi}{d\tilde{t}},\phi)$
determined by~\eqref{Ray1} and~\eqref{waveEq}, which, due to the
constraint~\eqref{Gauss1}, yield a dynamical system describing a flow on a
2-dimensional state space. Once the reduced system of first order
equations has been solved, equation~\eqref{Edota} yields
$\tilde{a} \propto \exp(\int d\tilde{t}\tilde{H})$.

Let us now follow the ideas presented in~\cite{alhugg15a,alhugg15} for scalar fields
and introduce new variables, which give a global dynamical system formulation
on the reduced Einstein state space:
\begin{subequations}\label{DepVar}
\begin{align}
\left(\tilde{T}, \tilde{X}, \Sigma_{\phi}\right) &= \left(\frac{M}{M + \tilde{H}},
\frac{M \left(1-e^{-\sqrt{\frac{2}{3}}\kappa\phi}\right)}{2\tilde{H}} ,
\frac{\kappa\frac{d\phi}{d\tilde{t}}}{\sqrt{6}\tilde{H}}\right), \\
\left(\tilde{H}, \kappa\phi, \kappa\frac{d\phi}{d\tilde{t}}\right) &=
\left(M\tilde{\ts}^{-1}, -\sqrt{3/2}\ln{\left(1-2\tilde{\ts}^{-1}\tilde{X}\right)} ,
\sqrt{6}M \tilde{\ts}^{-1}\Sigma_{\phi}\right),
\end{align}
\end{subequations}
where
\begin{equation}\label{TildeT_c}
\tilde{\ts} = \left(\frac{\tilde{T}}{1-\tilde{T}}\right),
\qquad M = 2\kappa\sqrt{\frac{V_0}{3}} = \frac{1}{\sqrt{6\alpha}},
\end{equation}
and a new time variable
\begin{equation}\label{bartaudef}
\frac{d\tilde{\tau}}{d\tilde{t}} = M\tilde{T}^{-1},
\end{equation}
which takes into account the different asymptotic scales of the
model, as described in~\cite{alhugg15a,alhugg15}.\footnote{In the present scalar
field context, this paper can be regarded as one in a series about scalar
field inflation models~\cite{alhugg15a,alhugg15}, and quintessence
models~\cite{ugg13,alhugg15c}, with the aim of showing how one can produce useful
dynamical systems and apply various dynamical systems and approximation techniques.}

This leads to the following evolution equations
\begin{subequations}\label{3Ddynsys}
\begin{align}
\frac{d\tilde{T}}{d\tilde{\tau}} &= 3\tilde{T}(1 - \tilde{T})^2 \Sigma^{2}_{\phi}, \\
\frac{d\tilde{X}}{d\tilde{\tau}} &=
\Sigma_{\phi}\left[3(1 - \tilde{T})\tilde{X}\Sigma_\phi + \tilde{T}F\right], \\
\frac{d\Sigma_{\phi}}{d\tilde{\tau}} &=
-\tilde{X}\left[3(1 - \tilde{T})\tilde{X}\Sigma_\phi + \tilde{T}F\right],
\end{align}
subjected to the constraint
\begin{equation}\label{EConstr}
1 = \Sigma^{2}_{\phi} + \tilde{X}^{2},
\end{equation}
\end{subequations}
where
\begin{equation}
\tilde{T}F = \tilde{T} - 2(1-\tilde{T})\tilde{X}.
\end{equation}
Note that, since
\begin{equation}
\frac{d(\tilde{T}F)}{d\tilde{\tau}} =
(1 - \tilde{T})\Sigma_\phi\left[3(1 - \tilde{T})\Sigma_\phi - 2\right](\tilde{T}F),
\end{equation}
$\tilde{T}F=0$ is an invariant boundary subset in the Einstein frame, but not in
the Jordan frame. This difference is due to the fact that the relation between the time variables in the two frames
is singular at $F=0$.

The relatively compact Einstein state space $\tilde{\bf S}$ is defined
by the cylinder with $0<\tilde{T}<1$ and with the region
$\tilde{T}F = \tilde{T} - 2(1-\tilde{T})\tilde{X} \leq 0$ cut out from it.
The state space can then be regularly extended to include the invariant boundary subsets
$\tilde{T}=1$, $\tilde{T}F = 0$, and $\tilde{T}=0$ when $\tilde{X}\leq 0$,
yielding the extended state space $\bar{\tilde{\bf S}}$ (see Figure~\ref{fig:Einsteinstatespace}).
Note that constant $\tilde{T}$ surfaces in the state space $\bar{\tilde{\bf S}}$ 
correspond to constant values of $\tilde{H}$, while the invariant boundaries are associated with
the asymptotic limits $\tilde{H}\rightarrow 0$ ($\tilde{T}=1$),   $\phi\rightarrow +\infty$ ($F=0$),
and $\tilde{H}\rightarrow \infty$ ($\tilde{T}=0$).
%

The constraint~\eqref{EConstr} can be globally solved by introducing
\begin{equation}\label{tildetheta}
\tilde{X}=\cos{\tilde{\theta}},\qquad \Sigma_{\phi}=\sin\tilde{\theta},
\end{equation}
which leads to the unconstrained 2-dimensional dynamical system
\begin{subequations}\label{2Ddynsys}
\begin{align}
\frac{d\tilde{T}}{d\tilde{\tau}} & = 3\tilde{T}(1 - \tilde{T})^2 \sin^{2}\tilde{\theta}, \label{tildeTevol}\\
\frac{d\tilde{\theta}}{d\tilde{\tau}} & =
\left(2 - 3\sin{\tilde{\theta}}\right)(1 - \tilde{T})\cos{\tilde{\theta}}- \tilde{T}.
\end{align}
\end{subequations}

Finally, the above changes of independent and dependent variables yield
$d\tilde{a}/d\tilde{\tau} = (1 - \tilde{T})\tilde{a}$, which leads to a
quadrature for $\tilde{a}$ once $\tilde{T}$ has been found. If one wants to express
the results in terms of the Einstein frame proper time variable $\tilde{t}$, one also needs
to integrate Eq.~\eqref{bartaudef}.

\subsection{Local fixed point analysis in the Einstein frame}\label{subsec:localein}

The dynamical system~\eqref{2Ddynsys} admits $4$ fixed points on $\bar{\tilde{\bf S}}$,
all located on the boundaries $\tilde{T}=0$ and $\tilde{T}F=0$:
\begin{subequations}
\begin{alignat}{8}
\mathrm{M}_{\pm}\!: \quad & \tilde{T} =\, & 0, \quad &\Sigma_{\phi} =\, & \pm 1, \quad & \tilde{X} = 0 &\quad
\rightarrow \quad & \tilde{\theta} =\,  2n\pi \pm \frac{\pi}{2},\\
\mathrm{PL}\!: \quad & \tilde{T} =\, & 0,
\quad &\Sigma_{\phi} =\, & \frac{2}{3}, \quad & \tilde{X} = -\frac{\sqrt{5}}{3} &\quad
\rightarrow \quad & \tilde{\theta} = \, \arccos\left(-\frac{\sqrt{5}}{3}\right) + 2n\pi,\\
\mathrm{dS}\!: \quad & \tilde{T} =\, & \frac{2}{3},
\quad &\Sigma_{\phi} =\, & 0, \quad & \tilde{X} =  1 &\quad
\rightarrow \quad & \tilde{\theta} =\,  2n\pi ,
\end{alignat}
\end{subequations}
where $n$ is an integer.

The Einstein frame deceleration parameter $\tilde{q}$, defined by
$d\tilde{H}/d\tilde{t} = -(1 + \tilde{q})\tilde{H}^2$, is given by
\begin{equation}\label{qscalar}
\tilde{q} = -1 + 3\Sigma^{2}_{\phi} = -1 + 3\sin^{2}\tilde{\theta}.
\end{equation}
It follows that $\mathrm{M}_{+}$ and $\mathrm{M}_{-}$ have $\tilde{q}=2$, which for the
minimally coupled scalar field interpretation corresponds to a massless state ($V(\phi)=0$),
while $\mathrm{dS}$ corresponds to a (quasi) de Sitter state in the Einstein frame
(associated with $\phi \rightarrow +\infty$),
since $\tilde{q} = -1$ for the solution that originates from this fixed point
asymptotically.\footnote{For further discussion on (quasi) de Sitter states,
see~\cite{alhugg15a,alhugg15}.} The notation $\mathrm{PL}$ stands for power law,
since the asymptotic behaviour of the solutions that originate from this fixed point in the
Einstein frame are described by the self-similar power-law solution associated with an exponential potential.

The fixed points $\mathrm{M}_{\pm}$ are hyperbolic sources with eigenvectors tangential to the
invariant subsets $\tilde{T}=0$ and $\tilde{T}F=0$. The fixed point $\mathrm{PL}$ is a hyperbolic saddle
with a single solution entering the state space ${\tilde{\bf S}}$. Finally the fixed point
$\mathrm{dS}$ has one negative eigenvalue and a zero eigenvalue, where the latter is associated with
the center manifold of $\mathrm{dS}$. The center manifold in turn corresponds to the inflationary attractor
solution in both the Einstein and Jordan state spaces, which is the single solution that enters
${\tilde{\bf S}}$ from $\mathrm{dS}$.

We now use center manifold analysis to establish that there is only a single solution that enters the state space
${\tilde{\bf S}}$ from $\mathrm{dS}$,\footnote{This was also established by means of center manifold analysis in~\cite{mir03},
although in other variables. Furthermore, together with~\cite{barher06}, which we will comment on further later on,
this paper gives the most complete description of the present models that we have found in the literature. Even so,
in contrast to the present paper, it does not give a global state space picture, and it is restricted to the
Einstein frame formulation.} and to obtain an approximation for this solution in the vicinity of
$\mathrm{dS}$ in ${\tilde{\bf S}}$. Linearizing the equations in the neighborhood of $\mathrm{dS}$ yields the following stable,
$E^s$, and center, $E^c$, tangential subspaces, respectively:
\begin{subequations}
\begin{align}
E^s &= \left\{(\tilde{T},\tilde{\theta})|\,\, \tilde{T} = \frac{2}{3}\right\},\\
E^c &= \left\{(\tilde{T},\tilde{\theta})|\,\,
\tilde{T}-\frac{2}{3} + \frac{\tilde{\theta}}{3} =0\right\}.
\end{align}
\end{subequations}
To investigate the center manifold $W^c$ associated with the tangent space
$E^c$, we adapt the variables to the location of $\mathrm{dS}$ and the tangent space
and replace $\tilde{T}$ and $\tilde{\theta}$ with
\begin{subequations}
\begin{align}
u &= \tilde{T} - \frac{2}{3}, \\
v &= \tilde{T} - \frac{2}{3} + \frac{\tilde{\theta}}{3},
\end{align}
\end{subequations}
so that $\mathrm{dS}$ is located at $(u,v) = (0,0)$.
%
The center manifold $W^c$ can be obtained as the graph $v = h(u)$ near $(u,v) = (0,0)$,
where $h(0) = 0$ (fixed point condition) and $\frac{dh}{du}(0) = 0$ (tangency condition).
Inserting these relationships into Eq.~\eqref{2Ddynsys} and using $u$ as the independent
variable leads to
\begin{equation}\label{tildehH}
\begin{split}
3 \tilde{T}(u) (1-\tilde{T}(u))^2 &\sin^{2}\tilde{\theta}(u)\left(\frac{d h(u)}{du} - 1\right) \\
& -\frac{1}{3}\left[\left(2-3\sin{\tilde{\theta}(u)}\right)
(1-\tilde{T}(u))\cos{\tilde{\theta}(u)}- \tilde{T}(u)\right] = 0,
\end{split}
\end{equation}
%
%
where $\tilde{T}(u)= u + 2/3$ and $\tilde{\theta}(u)=3(h(u)-u)$.
As before, we can solve the equation approximately by representing $h(u)$ as a formal power
series
%
truncated at some chosen order $n$. Inserting
this into Eq.~\eqref{tildehH} and algebraically solving for the coefficients leads to
\begin{equation}
h(u) = -2u^2 - 6u^3 + {\cal O}(u^4).
\end{equation}
%
%
%
It follows that the single solution that originates from $\mathrm{dS}$ into ${\tilde{\bf S}}$
(the `inflationary attractor solution') is described by the approximate expansion
\begin{equation}\label{ThetaExpu}
\tilde{\theta}(u) = -3u\left\{1 + 2u + 6u^2 + {\cal O}(u^3)\right\}.
\end{equation}
%

\subsection{Global analysis in the Einstein frame}\label{subsec:globalein2}

In this case $\tilde{H}$ is monotonically decreasing, except when $\tilde{q}=-1$, which corresponds
to the following monotonicity properties of $\tilde{T}$ in our Einstein frame state space setting.
From its evolution equation~\eqref{tildeTevol} we see that $\tilde{T}$ is monotonically increasing
in $\tilde{\bf S}$ when $\tilde{q}\neq -1$. Since by~\eqref{qscalar} $\tilde{q}=-1$ corresponds to
$\tilde{\theta}=2n\pi$, we have
\begin{equation}\label{mon}
\left. \frac{d\tilde{T}}{d\tilde{\tau}}\right|_{\tilde{\theta}=2n\pi} = 0,\qquad
\left. \frac{d^2\tilde{T}}{d\tilde{\tau}^2}\right|_{\tilde{\theta}=2n\pi} = 0,\qquad
\left. \frac{d^3\tilde{T}}{d\tilde{\tau}^3}\right|_{\tilde{\theta}=2n\pi} =
18\tilde{T}(1-\tilde{T})^2(\frac{2}{3} - \tilde{T}),
\end{equation}
and thus $\tilde{q} = -1$ only represents an inflection point in the evolution of
$\tilde{T}$ in $\tilde{\bf S}$ (there are no invariant sets at $\tilde{T} = 2/3$ in $\tilde{\bf S}$).
The monotonicity of $\tilde{T}$ in combination with the expression
for $d^3\tilde{T}/d\tilde{\tau}^3|_{\tilde{\theta}=2n\pi}$, shows that solutions in $\tilde{\bf S}$
either come from $\tilde{T} = 0$ or from $\mathrm{dS}$. Combining this with the previous
local analysis show that there are two one-parameter sets of solutions entering $\tilde{\bf S}$
from $\mathrm{M}_+$ and $\mathrm{M}_-$, respectively, a single solution entering from $\mathrm{PL}$,
and one from $\mathrm{dS}$.

\begin{figure}[ht!]
\centering
\subfigure[The state space of the Einstein frame.]{\label{fig:EinsteinSScylinder}
\includegraphics[width=0.45\textwidth, trim = 0cm 0.25cm 0cm 0cm]{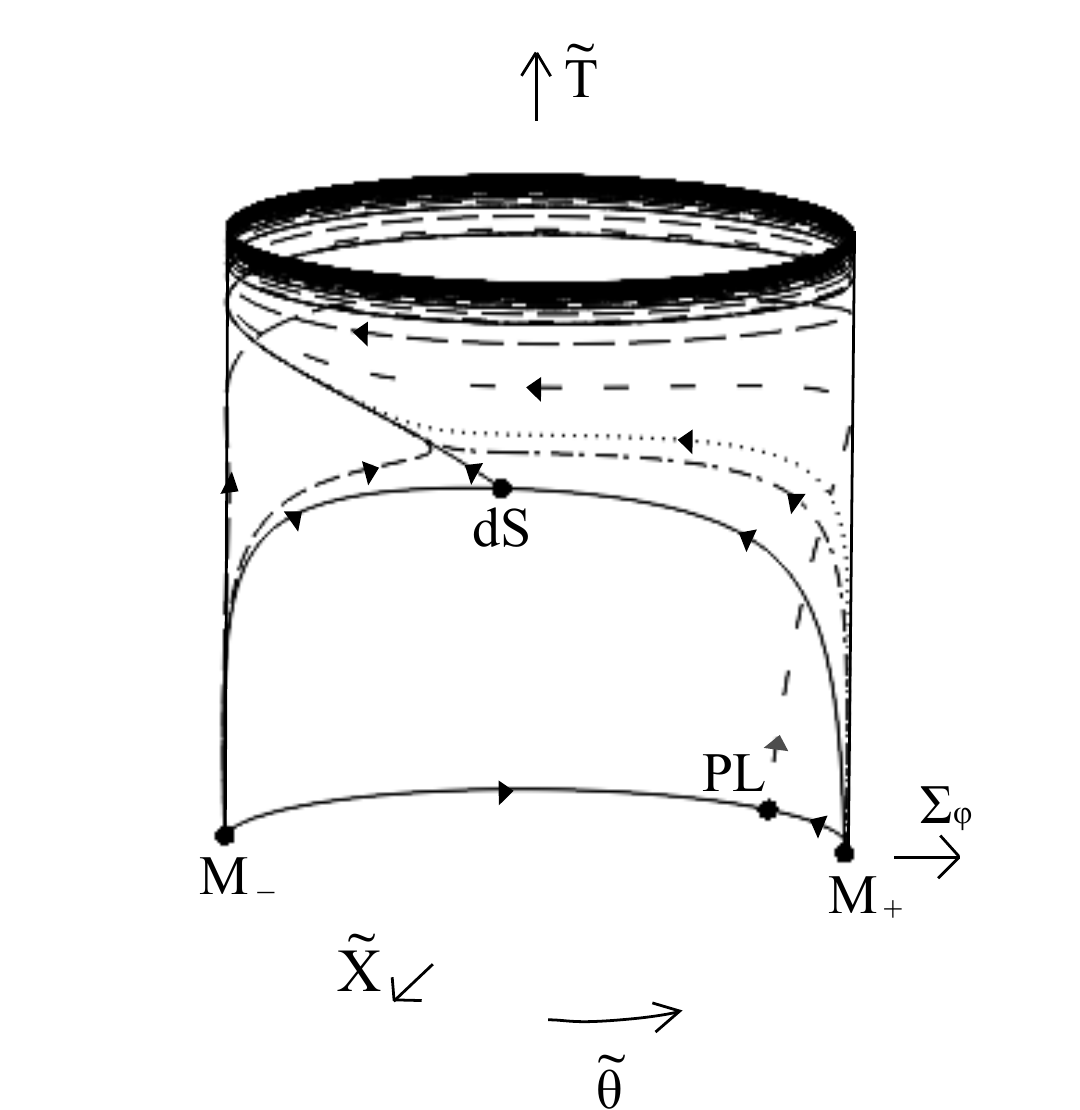}}
\subfigure[The unwrapped state space of the Einstein frame.]{\label{fig:EinsteinWraped}
\includegraphics[width=0.40\textwidth]{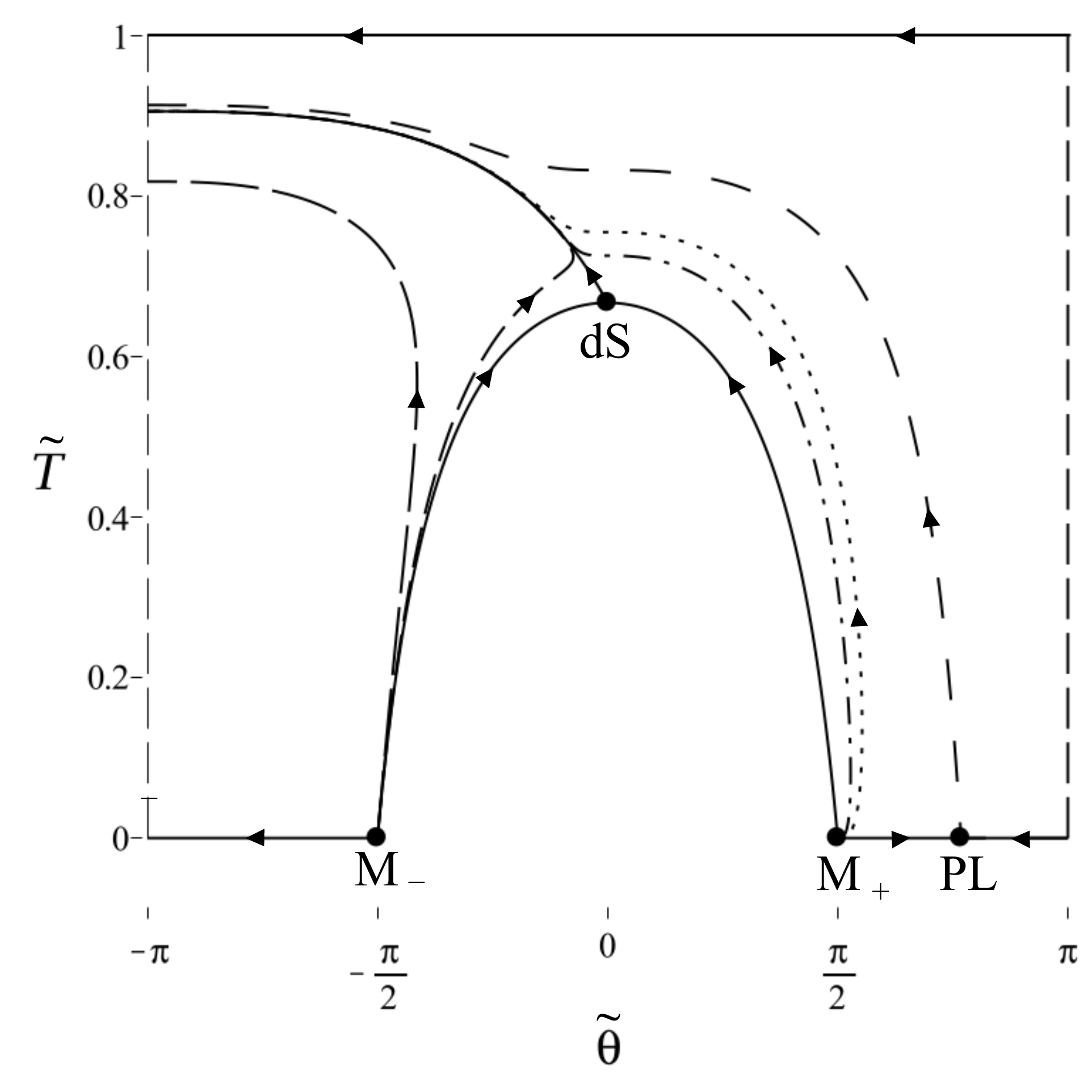}}
\caption{The extended state space $\bar{\tilde{\bf S}}$ of the Einstein frame, consisting of a
finite cylinder with a removed non-physical $\tilde{T}F<0$ region, and representative solutions.
}
\label{fig:Einsteinstatespace}
\end{figure}

At the boundary subset $\tilde{T} = 1$ we obtain
\begin{equation}
\frac{d\tilde{\theta}}{d\tilde{\tau}}  = - 1
\end{equation}
from~\eqref{2Ddynsys}. This shows that $\tilde{T}=1$ is a periodic orbit with
monotonically decreasing $\tilde{\theta}$. Furthermore, the monotonicity of $\tilde{T}$
shows that this is the limit cycle of all orbits in $\tilde{\bf S}$, thus
constituting their $\omega$-limit set. The solution space in the Einstein frame is depicted in
Figure~\ref{fig:Einsteinstatespace}.

\begin{figure}[ht!]
\begin{center}
{\subfigure[The averaged solution at late times.]{\label{fig:average}
\includegraphics[width=0.45\textwidth]{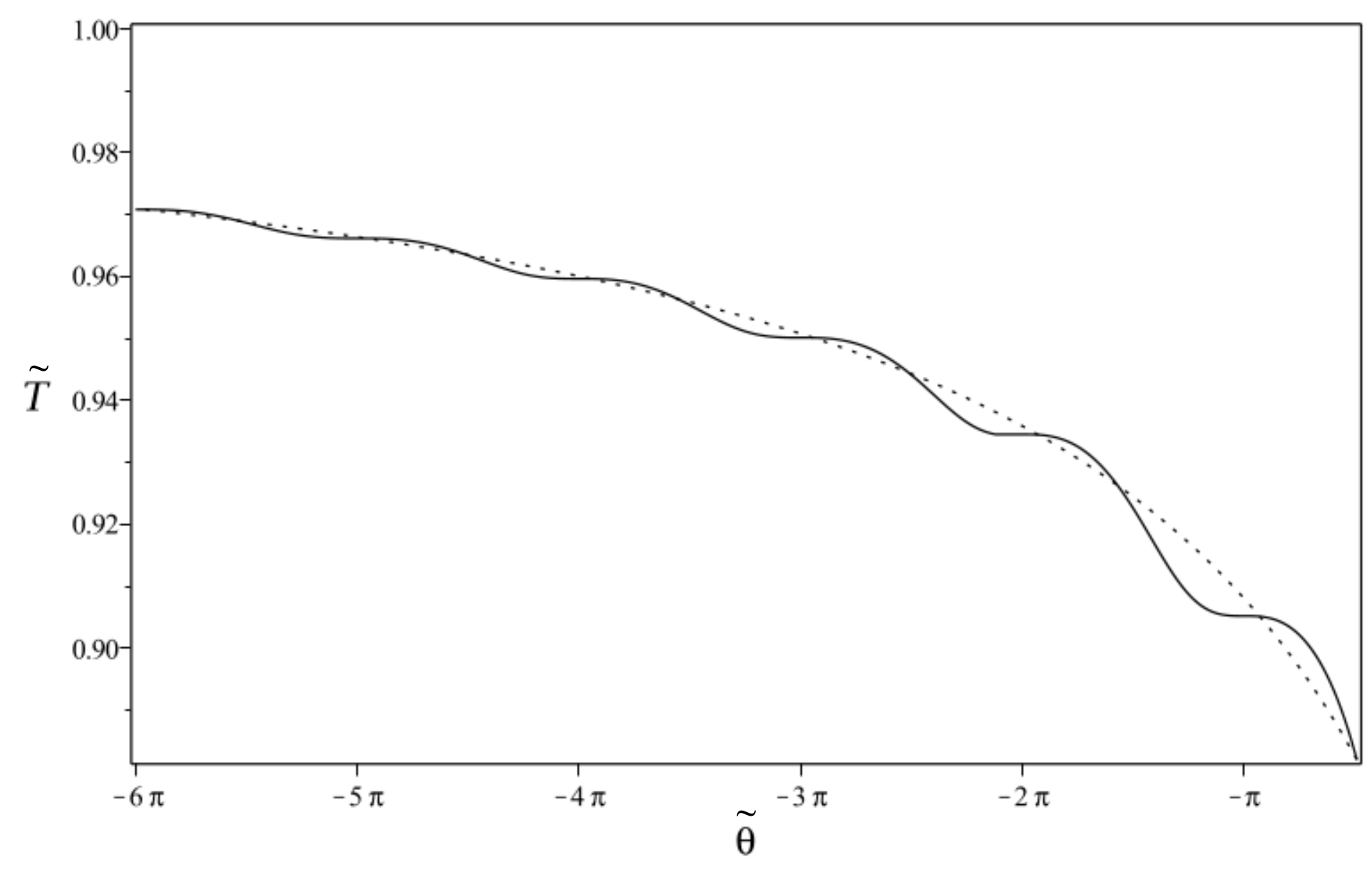}}}\qquad
{\subfigure[An oscillatory approximation for the oscillatory late time regime]{\label{fig:oscback}
\includegraphics[width=0.45\textwidth]{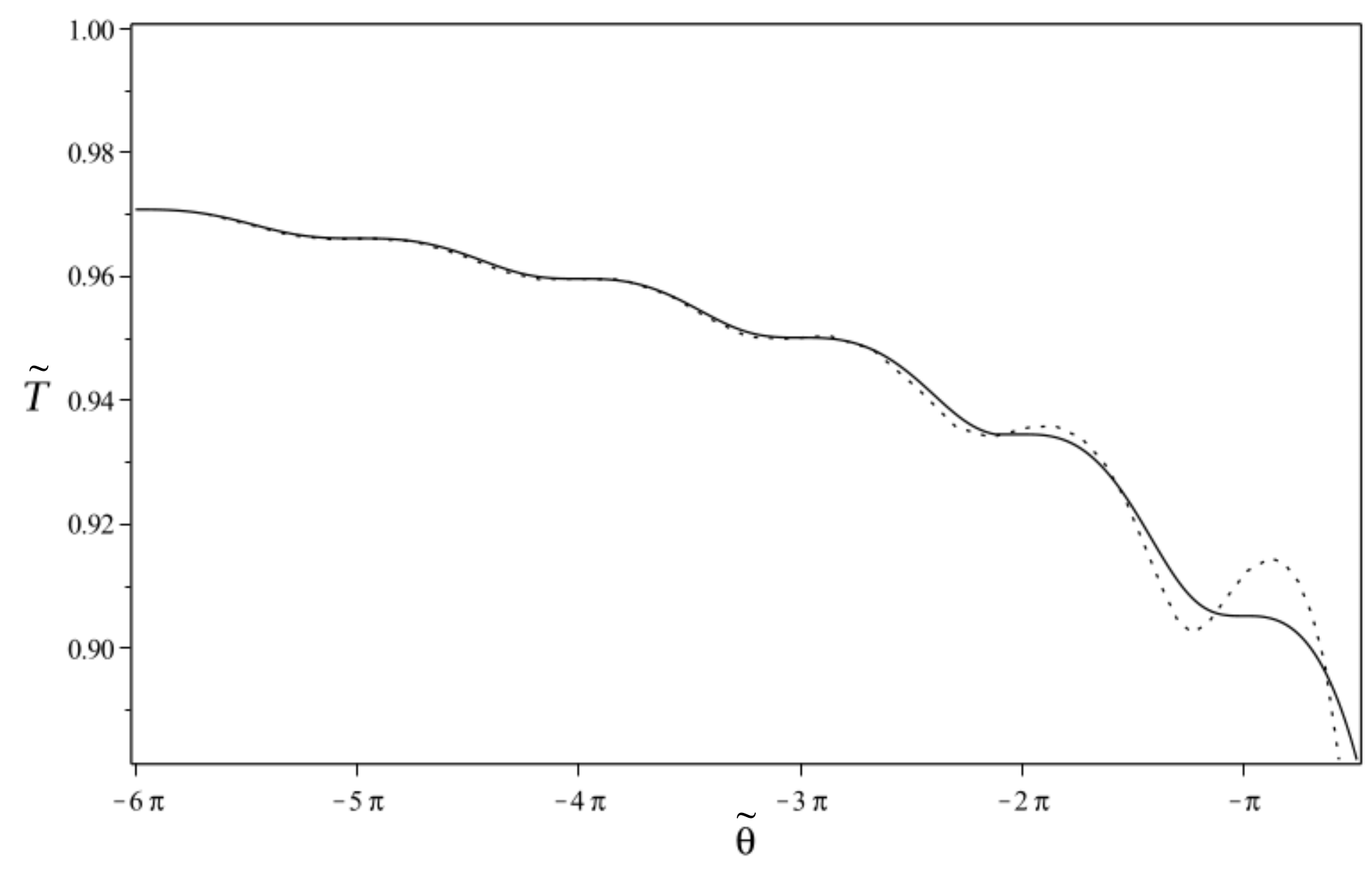}}}
\end{center}
\caption{The plots show the behavior at late times of the numerically computed
inflationary attractor solution (solid line),
the averaged solution, and the oscillatory late time approximation
in the Einstein frame state space (the dotted lines in the two figures).
}
\label{fig:latetimes}
\end{figure}

We now present some approximation schemes for solutions at late times close to $\tilde{T}=1$, which corresponds to
$\tilde{H}\rightarrow 0$, i.e., we will give approximations for the oscillatory phase at late times.
We first consider an averaging technique
used in~\cite{alhugg15a}. We thereby take the average with respect to $\tilde{\theta}$ of the right
hand side of~\eqref{2Ddynsys} (since $-\tilde{\theta} \rightarrow \tilde{\tau} \propto \tilde{t}
\rightarrow \infty$ while $\tilde{T}$ slowly approaches one), which leads to
\begin{subequations}\label{dynsysaverage}
\begin{align}
\frac{d\tilde{T}}{d\tilde{\tau}} &= \frac32 \tilde{T}(1-\tilde{T})^2,\label{Teqav}\\
\frac{d\tilde{\theta}}{d\tilde{\tau}} &= - \tilde{T}  .
\end{align}
\end{subequations}
It follows that
\begin{equation}\label{thetaTlate}
\tilde{\theta} = - \frac{2}{3(1-\tilde{T})} + C,
\end{equation}
where $C = \tilde{\theta}_i + 2/3(1-\tilde{T}_i)$, where $(\tilde{\theta}_i, \tilde{T}_i)$ is some initial
point for the trajectory. This approximation is valid for all
solutions when $\tilde{T}$ approaches one, including the center manifold
attractor solution, see Figure~\ref{fig:average}.

In~\cite{ren07} Rendall gave rigorous results for late time behavior for scalar field models
with a potential with a minimum that asymptotically can be described by a $\phi^2$ potential.
Since this covers the present models, we can translate the results in~\cite{ren07},
which yields the following asymptotic approximation:
\begin{equation}\label{thetaTlateosc}
\tilde{\theta} = -\tilde{t} - \frac{3+2\cos(2\tilde{t})}{4\tilde{t}}, \qquad
\tilde{T} = \left(1 + \frac{2}{3(\tilde{t}-\tilde{t}_0)}\left(1 + \frac{\sin(2\tilde{t})}{2\tilde{t}}\right)\right)^{-1},
\end{equation}
where $\tilde{t}_0$ is a constant and $\tilde{t}$ is proper time in the Einstein frame
(with $M$ normalized to one). The relations given in Eq.~\eqref{thetaTlateosc} describe a
parameterized curve in the global state space ${\tilde{\bf S}}$, which is plotted in
Figure~\ref{fig:oscback}; note that the oscillatory approximation becomes
increasingly accurate toward the future, reflecting that it describes the
asymptotic evolution at late times.
%

\subsection{Situating the Einstein frame state space in that of the Jordan frame}

\begin{figure}[ht!]
\begin{center}
\includegraphics[width=0.5\textwidth]{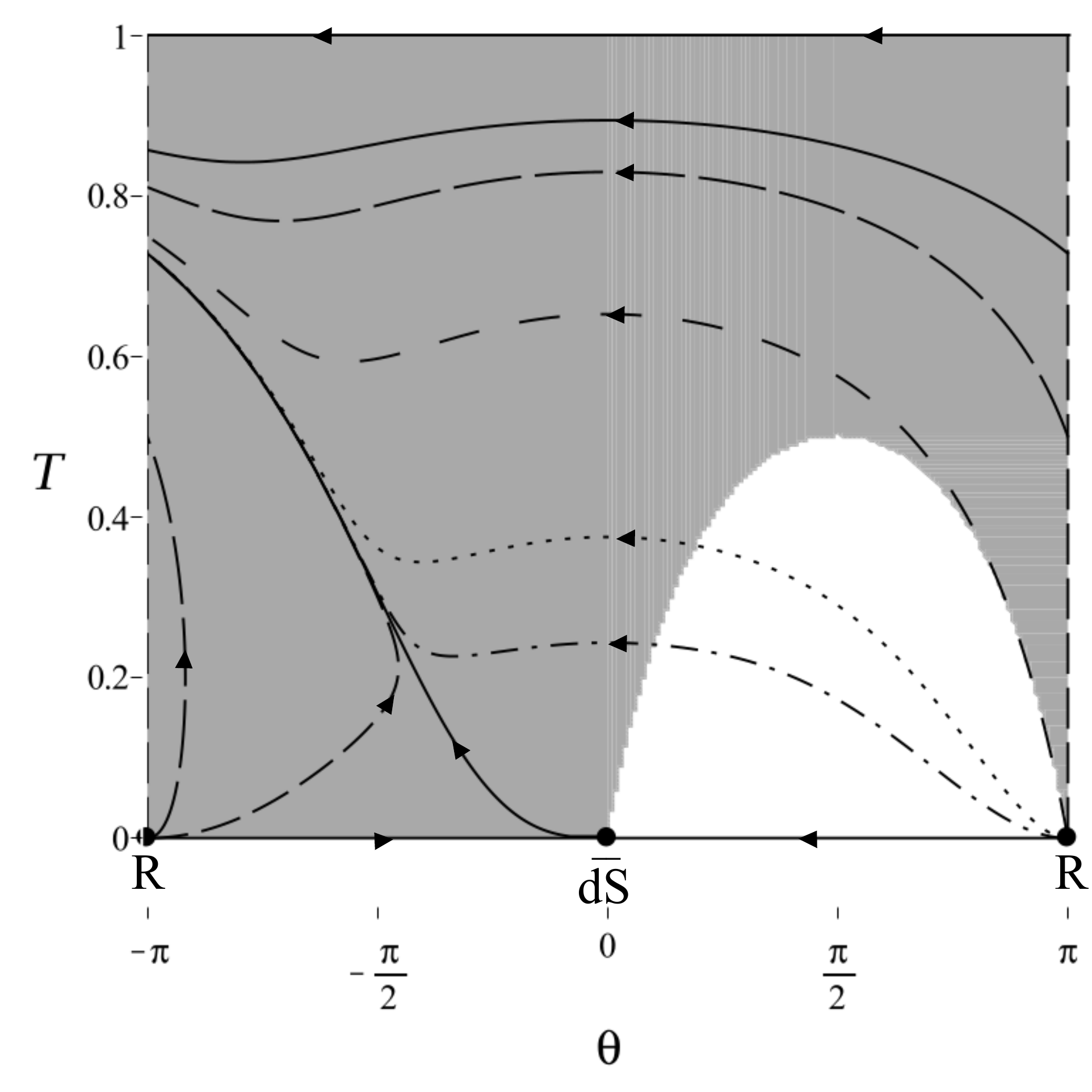}
\end{center}
\caption{As in Figure~\ref{fig:LightCone_SS}, the shaded region in the figure above depicts the domain of the Jordan state space that is conformal
to the Einstein frame, given by $F=1+2\alpha R>0$. Note that the special space-dashed solution, originating from $\mathrm{R}$ in
the direction transverse to the invariant boundary $T=0$, divides the solutions into those that enter
$\mathbf{S}$ from the $F<0$ region (exemplified by the dotted and  dash-dotted lines) and those
that are always in the region $F>0$ (exemplified by the dashed and long-dashed lines and the
inflationary attractor solution originating from $\bar{\mathrm{dS}}$).
}
\label{fig:Jordan_EinsteinBound}
\end{figure}

The Einstein frame is characterized by the conformal factor $F$ which has to satisfy $F>0$,
and therefore the Einstein frame state space is characterized by a boundary $F=0$ in the
Jordan frame state space. In the present case the conformal factor $F=1+2\alpha R$ is given in terms of
our Jordan frame state space variables by
\begin{equation}
F(\theta,T)=1-\frac{(1-T)}{T}\sin{\theta}
\end{equation}
The curve $F=0$ intersects the invariant boundary $T=0$ at $\theta=2n\pi$ and $\theta=\pi+2n\pi$, i.e.,
at the fixed points $\bar{\mathrm{dS}}$ and $\mathrm{R}$. The variable $T$ has a maximum along the curve
$F=0$ at $T=1/2$, which happens when $\theta=\pi/2+2n\pi$.
Moreover, at $F=0$ it follows that 
\begin{equation}
\left. \frac{dF}{d\bar{t}}\right|_{F=0}=2T(1+\cos{\theta})
\end{equation}
which is everywhere positive except at $\theta=\pi+2n\pi$, i.e., at the fixed point $\mathrm{R}$ on
the boundary of the Jordan frame state space where it vanishes (see Figure~\ref{fig:Jordan_EinsteinBound}).
This implies that there are no orbits that pass through $F=0$ from the region $F>0$ into the state
space where $F<0$.

The relation between the solutions originating from the fixed points in the Jordan and Einstein frame
state spaces can be obtained by noting that
\begin{equation}\label{qevar}
q = 1- \left(\frac{\tilde{T}}{1 - \tilde{T}}\right)\frac{\tilde{X}}{(1- \Sigma_\phi)^2}.
\end{equation}
By inserting the fixed point $\mathrm{dS}$ into the above expression, it follows that
the asymptotic (quasi) de Sitter state $\mathrm{dS}$ in the Einstein frame state space is also the
asymptotic (quasi) de Sitter state in the Jordan frame state space. Thus $\tilde{H}$ is asymptotically
finite for the inflationary attractor solution in the Einstein frame, while $H\rightarrow \infty$
in the Jordan frame.


We also note that $q=1$ for $\mathrm{PL}$ and $\mathrm{M}_-$, while the above expression for $q$ is ill-defined
for $\mathrm{M}_+$. However, since $\mathrm{M}_+$ is hyperbolic and  the dynamical
system~\eqref{2Ddynsys} is analytic we can insert the solution of the linearized equations in the neighbourhood
of $\mathrm{M}_+$ to find a suitable approximation for $q$. It turns out that near $\mathrm{M}_+$ $q$
becomes a constant with the value determined by the ratio of the two arbitrary
constants associated with the two eigenvectors. This describes a 1-parameter set of solutions
passing through $F=0$ from negative to positive $F$ in the Jordan frame state space.
These results establish that $\mathrm{PL}$ and $\mathrm{M}_-$ yield the solutions
that originate from the fixed point $\mathrm{R}$ in the Jordan state space (where the solution from
$\mathrm{PL}$ corresponds to the solution that initially is tangential to $F=0$ in the Jordan frame state
space) while the solutions that originate from $\mathrm{M}_+$ correspond to a coordinate
singularity associated with the breakdown of the Einstein frame at $F=0$.

The solution that originates from $\mathrm{PL}$ divides the orbits that
originate from $\mathrm{R}$ in the Jordan frame into two classes: (a) Orbits, like itself and the single orbit
from $\bar{\mathrm{dS}}$ (which is the same orbit as that coming from $\mathrm{dS}$ in the Einstein frame),
that throughout their evolution have $F>0$ (i.e. their evolution in the Jordan frame is entirely covered by
that in the Einstein frame), and (b) orbits that begin with $F<0$ in the Jordan frame and
then pass through $F=0$ and subsequently have $F>0$ throughout their remaining evolution. The last class,
therefore, consists of solutions in the Einstein frame that are past conformally extendible in the Jordan frame; see
Figure~\ref{fig:Jordan_EinsteinBound} where the shaded region corresponds to the region in the Jordan state space
that is conformal to the Einstein frame (cf. Figure~\ref{fig:LightCone_SS}; note that is easy to translate the
present results to the original state space picture by means of the figures). If one is so inclined, one can obtain
further details of the above nature by inserting approximate asymptotic solutions into
Eq.~\eqref{JorEinvar} in Appendix~\ref{app:relations}, which describes the transition between the Jordan and
Einstein frame state space variables.
%

\section{Discussion}\label{sec:disc}

The global regular system we have presented for the Jordan frame naturally conveys
the global properties of the models at hand, as illustrated by Figure~\ref{fig:VacuumJF}.
Nevertheless, it is by no means an optimal dynamical system for all aspects one might
want to investigate: there are other dynamical systems that have complementary properties.
This is already exemplified by the fact that our dynamical system for the Einstein frame,
among other things, simplifies the asymptotic analysis of the inflationary
attractor solution originating from the $\mathrm{dS}$ fixed point and offers various approximation 
schemes for the oscillatory regime at late times, thereby complementing other 
heuristic Jordan frame methods~\cite{mijetal86,LivRev10}. Another
useful system is discussed in Appendix~\ref{app:systems} (where its close relationship
to the works in~\cite{barher06,car15} is also commented on). It is based on a
variable transformation from $(H, R)$ to the variables $(z,q)$, defined by
\begin{equation}\label{Defzq}
z = \frac{1}{12\alpha H^2}, \qquad q = 1- \frac{R}{6H^2},
\end{equation}
and the time variable $N=\ln a$,
which leads to the following simple regular system of unconstrained equations:
\begin{subequations}\label{zqsys}
\begin{align}
\frac{dz}{dN} &= 2(1+q)z,\\
\frac{dq}{dN} &= z - \frac32(1-q^2).
\end{align}
\end{subequations}

As discussed in Appendix~\ref{app:systems}, the variable transformation breaks down at $H=0$, which for $z$  and $q$
are located at infinity, i.e., $z$ and $q$ are unbounded. Straightforward compactifications such as a Poincar{\'e}
compactification of $z$ and $q$ are inappropriate, since such compactifications result in an erroneous state space
topology, which may result in wrong conclusions about the properties of the solutions. For example, such a
compactification ruins, or at least complicates, a treatment of the oscillatory regime at late times.
\emph{This illustrates that it is necessary to take into account the global topological
properties of the physical state space in order to obtain a correct description of the solution space and its
properties}, which illustrates a non-local aspect (apart from fixed points reflecting coordinate singularities) concerning
the relationship between dynamical systems formulations and (iv): physical solution space interpretation.
Nevertheless, the above system has local advantages. The system
admits two fixed points, both located on the invariant boundary  $z=0$ ($H\rightarrow+\infty$):
\begin{subequations}\label{FP_qz}
\begin{alignat}{2}
\mathrm{R}\!:\quad z &=0, &\qquad q &= 1, \\
\mathrm{dS}\!: \quad z &=0, &\qquad q &= -1.
\end{alignat}
\end{subequations}
The fixed point $\mathrm{R}$ is a hyperbolic source and corresponds to $\mathrm{R}$ in our global
system, while $\mathrm{dS}$ is non-hyperbolic with one negative eigenvalue and one zero eigenvalue. 
This fixed point corresponds to the fixed point $\mathrm{dS}$ in the global system, obtained
after a blow up. A center manifold analysis associated with the zero eigenvalue of $\mathrm{dS}$
yields the following approximation for the inflationary attractor solution (see Appendix~\ref{app:systems}):
\begin{equation}\label{qzcenter}
q(z) = -1 + \frac{z}{3}\left[1 + \frac{z}{3} - \frac{z^3}{3^5} + \dots\right].
\end{equation}
Note that obtaining an approximation for the inflationary attractor solution that comes from $q=-1$ when
$H\rightarrow \infty$ is considerably easier in these variables than for the $T$, $\theta$ variables,
and the expansion is given as $q(H^{-2})$ since $z \propto H^{-2}$, which might be regarded as
preferable.

The above brings the inflationary attractor solution into focus. Usually the inflationary
regime is understood in terms of slow-roll approximations. In the Jordan frame this approximation
can be found in~\cite{mijetal86,LivRev10} and reads $\dot{H} = -1/36\alpha$.
Since $\dot{H} = -(1+q)H^2$, this leads to $1+q= 1/36\alpha H^2$, which gives $q = -1 + z/3$,
i.e., it just gives the leading order term in the center manifold expansion~\eqref{qzcenter}
for the inflationary attractor solution in the variables $z,q$.
A comparison with the global variables shows that to leading order $z = (\theta/2)^2$.
The above illustrates that not only are the $z,q$ variables a useful complement to the global
variables $T,\theta$, since they more straightforwardly give approximations for the inflationary
regime, but they are also intimately linked to the Hubble slow-roll approach.

Next we consider the Einstein frame  and the usual slow-roll approximation. In this setting the
slow-roll approximation is obtained by inserting $\tilde{H}=\kappa\sqrt{V(\phi)/3}$ into
\begin{equation}\label{dotphiHphi}
\kappa\frac{d\phi}{d\tilde{t}} = - 2\frac{\partial \tilde{H}}{\partial \phi},
\end{equation}
which for the present scalar field potential gives
\begin{equation}\label{SR}
\kappa\frac{d\phi}{d\tilde{t}} \approx -\sqrt{\frac{2}{3}} M e^{-\sqrt{\frac{2}{3}}\kappa\phi} .
\end{equation}
Expressed in terms of the variables $\Sigma_\phi$ and $\tilde{X}$, this
results in
\begin{equation}\label{slow-roll}
\Sigma_\phi \approx -\frac{1}{3} \left[\left(\frac{\tilde{T}}{1-\tilde{T}}\right) - 2\tilde{X}\right].
\end{equation}
In the neighborhood of $\mathrm{dS}$, represented by $\tilde{\theta}=\tilde{T}=0$, this yields
\begin{equation}
\tilde{\theta} \approx -3\left(\tilde{T}-\frac{2}{3}\right),
\end{equation}
which is the tangency condition for the center submanifold of $\mathrm{dS}$, given by the leading order
expression in~\eqref{ThetaExpu}. The slow-roll approximation is therefore just an approximation for the center
manifold in the vicinity of $\mathrm{dS}$ in our Einstein frame state space formulation.
In this context it should be pointed out that we can of course use variable relationships, given in Eq.~\eqref{JorEinvar} in
Appendix~\ref{app:relations}, to translate the various approximations from the Einstein to the Jordan frames and vice versa,
and their series expansions can be improved by taking Pad{\'e} approximants, as discussed in e.g.~\cite{alhugg15a}.

We end this discussion by emphasizing once more that the main purpose of the presently studied models was to specifically illustrate
some general aspects of $f(R)$ cosmology with a simple example, namely (a) the ingredients (i) -- (iv) in the introduction,
and (b) some dynamical systems methods with a broad range of applicability, even though the particulars
have been tailored to the specific properties of the $f=R + \alpha R^2$ RW models. Although quite special, these
models also capture very clearly some central issues in $f(R)$-gravity beyond the above methodological
aspects. For example, as stated in the introduction, one way of classifying $f(R)$-gravity models is according to if $F>0$ for
all $R$ or not. In the latter case the correspondence between the original $f(R)$ model and its Einstein frame formulation,
or its Brans-Dicke ($\omega_\mathrm{BD}=0$) version (see e.g.~\cite{LivRev10,aveetal16} for a description of this
correspondence), only holds locally for the range of $R$ where $F>0$. For such models the evolution in the Jordan
frame of some solutions are incompletely described in these formulations, i.e., a local formulation correspondence does
not entail a global correspondence, as is clearly illustrated in Figure~\ref{fig:Jordan_EinsteinBound}.

In this context, note that if $F=0$ was an invariant subset in the Jordan frame, then the $F<0$ part of the
Jordan state space would constitute an invariant subset. In this case, one could perhaps argue that the solutions associated with
this part of the state space could be discarded on some claimed physical grounds, thus leading to a global physical
correspondence between the solution spaces of the different frame formulations. However, as we have shown,
$F=0$ is not in general an invariant subset, nor is therefore $F<0$. Thus if one wants to argue that a global physical
correspondence exists for the different formulations, one is forced to come up with some arguments for why part of some solutions
in the Jordan frame should be discarded (note that the existence of such solutions is ensured by that $F=0$ is not an invariant
subset; again, see Figure~\ref{fig:Jordan_EinsteinBound} as an illustrative example).

There are, of course, some things that the present models cannot address. In particular this holds for models where
the condition $F_{,R}>0$ is broken (leading to e.g. tachyonic instabilities, see e.g.~\cite{LivRev10} and
references therein). The change of sign of $F_{,R}$ is particularly problematic from a mathematical point of view
since the constraint~\eqref{constJorf} becomes degenerate when $F_{,R}=0$, and, moreover, the causal properties of
the field equations change when $F_{,R}$ changes sign (hence the tachyonic instability). There has been some
work in $f(R)$ cosmology to extend solutions when the equations are ill-defined, notably~\cite{barher06}.
However, we here point out that the existence of
$F_{,R}=0$ state space boundaries mathematically resemble sonic shock waves for fluids. It is therefore worthwhile
to note that such problems have been dealt with in e.g. the context of spherically symmetric self-similar perfect fluid
models~\cite{goletal98a}--\cite{caretal01}, where it was shown how to extend solutions through sonic
shock wave surfaces. Incidentally, these models also provide examples where it is useful to cover the state space with several
coordinate patches in order to exploit special structures in different parts of the state space, a problem one will inevitably
will have to deal with when it comes to most $f(R)$ cosmological models.

\subsection*{Acknowledgments}
A. A. is funded by the FCT grant SFRH/BPD/85194/2012, and supported by the project PTDC/MAT-ANA/1275/2014,
and CAMGSD, Instituto Superior T{\'e}cnico by FCT/Portugal through UID/MAT/04459/2013.
S. C.  was supported by  the Funda\c{c}\~{a}o para a Ci\^{e}ncia e Tecnologia through project IF/00250/2013.
S. C. acknowledges financial support provided under the European Union's H2020 ERC Consolidator Grant
``Matter and strong-field gravity: New frontiers in Einstein's theory'' grant agreement no.  MaGRaTh646597,
and under the H2020-MSCA-RISE-2015 Grant No. StronGrHEP-690904. C. U. thank the Center for Mathematical Analysis,
Geometry and Dynamical Systems, Instituto Superior T\'ecnico, Lisbon, for kind hospitality.

\begin{appendix}

\section{Jordan and Einstein state space relationships}\label{app:relations}

To translate between our global dynamical systems formulations for the Jordan and Einstein
frame state spaces, we need the explicit relationship between our global dynamical systems variables,
which are as follows:
%
\begin{subequations}\label{JorEinvar}
\begin{xalignat}{2}
H &= M\left(\frac{1-\tilde{T}}{\tilde{T}}\right)(1 - \Sigma_\phi)F^{-1/2}, &\quad
2\alpha\ts &= \frac{1}{\sqrt{2}}\left(\frac{1-\tilde{T}}{\tilde{T}}\right)F^{-3/2}G_+, \\
\dot{R} &= \frac{M}{\alpha}\left(\frac{1-\tilde{T}}{\tilde{T}}\right)\Sigma_\phi F^{-3/2}, &\quad
X &= G_-G_+^{-1}, \\
R &= \frac{1}{\alpha}\left(\frac{1-\tilde{T}}{\tilde{T}}\right)\tilde{X}F^{-1}, &\quad
S &= -2\sqrt{2}\tilde{X}F^{1/2}G_+^{-1},
\end{xalignat}
and
\begin{xalignat}{2}
\tilde{H} &=\frac{M}{4\sqrt{2}}\left(\frac{1-T}{T}\right)H_+ F^{-3/2} , &\quad
\tilde{\ts} &= 4\sqrt{2}\left(\frac{T}{1-T}\right)F^{3/2}H_{+}^{-1}, \\
\kappa\frac{d\phi}{d\tilde{t}} &=\frac{1}{4\sqrt{2\alpha}}\left(\frac{1-T}{T}\right)H_-F^{-3/2}, &\quad
\Sigma_{\phi} &= H_-H_+^{-1},\\
\kappa\phi &=\sqrt{\frac{3}{2}}\ln{F}, &\quad
\tilde{X} &=-2\sqrt{2}SF^{1/2}H_+^{-1},
\end{xalignat}
\end{subequations}
where we have defined
\begin{subequations}
\begin{align}
G_\pm  &= 1 + \Sigma_\phi \pm 2(1-\Sigma_\phi)F, \\
H_{\pm} &= 2(1+X)\pm(1-X)F,
\end{align}
\end{subequations}
and where we recall that
\begin{equation}
\begin{split}
M &= \frac{1}{\sqrt{6\alpha}}, \qquad \tilde{T}=\frac{\tilde{\ts}}{1+\tilde{\ts}},  \qquad T = \frac{1}{1 + 2\alpha\ts}, \\
F &= 1 - 2\left(\frac{1-\tilde{T}}{\tilde{T}}\right)\tilde{X}=1-\frac{(1-T)}{T}S.
\end{split}
\end{equation}
As expected it follows that the variable transformation from the future light cone Jordan state space
to the Einstein state space breaks down at the boundary $\tilde{T}F=0$, since the Jacobian determinant for
the variable transformation $(H,\dot{R},R) \rightarrow (\tilde{T},\Sigma_\phi,\tilde{X})$ is given by
$-(1-\tilde{T})^2(\tilde{T}F)^{-4}/6\alpha^3$.


\section{Various dynamical systems formulations}\label{app:systems}

Consider the dynamical systems formulation in the Jordan frame based on the variable
transformation from $(H, R)$ to the variables $(z,q)$, defined in~\eqref{Defzq} and
which obey the evolution equation given in \eqref{zqsys},
%
%
while the constraint is used to solve for $\dot{R}$.
%
%
This system is manifestly invariant under the discrete symmetry $(t,H) \rightarrow -(t,H)$, and it is remarkably
simple. So why not use this system of equations instead of the previous ones?\footnote{The
present variables $z$ and $q$ are closely related to those in~\cite{barher06} and~\cite{car15}.
Indeed, they are affinely related to $B$ and $Q$ in~\cite{barher06}, while $z$ is proportional to
$\mathbb{A}^{-1}$ and $q$ is affinely related to $\mathbb{Q}$ in~\cite{car15}.
}

Firstly, note that the Jacobian determinant of the variable transformation $(z,q)$ to $(H, R)$ is given by
$1/(36\alpha H^5)$, i.e., the variable transformation breaks down at $H=0$, which for $(z,q)$ is located
at infinity. Further insight is obtained by using our definitions to express $z$ and $q$ in terms of
$T$ and $\theta$:
\begin{equation}
z = \left[\left(\frac{T}{1-T}\right)\!\left(\frac{1+\tan^2(\theta/2)}{\tan^2(\theta/2)}\right)\right]^2, \quad
q=1+2\left(\frac{T}{1-T}\right)\!\left(\frac{1+\tan^2(\theta/2)}{\tan^3(\theta/2)}\right).
\end{equation}
As can be seen, both $z$ and $q$ diverges when $\theta = 2n\pi$, i.e., when $H=0$, which is where the
variables break down. Furthermore, all solutions pass through $H=0$ (where $q$ blows up) infinitely many times
during the oscillating era at late times. Our Jordan state space formulation in the main text
has also the advantage of clearly showing that $q \rightarrow \pm \infty$ is not associated with any 
spacetime singularity, but instead reflects the fact that $H$ becomes zero during the cosmic evolution.



Secondly, all the present variables are unbounded, although $z$ is
positive. In addition, it is possible to use the natural extension of the state space that includes the $z=0$ invariant subset 
boundary, since all interior orbits originate from fixed points on this boundary,
which corresponds to $H\rightarrow \infty$. One can then use that $z=0$ is an invariant boundary \emph{and} that $z$
is non-negative to produce a new bounded variable, defined by $z/(1+z)$, and change the time variable
appropriately so that the right hand sides of the equations become polynomial in
the dependent variables. However, a similar procedure is not possible with $q$. Of course one can
replace $q$ with a bounded variable in a number of ways, but there does not seem to exist a physical
structure which one can tie to such a compactification, except by essentially going back to our original bounded system.
For example, a Poincar{\'e} compactification of $z,q$ would enforce an erroneous topology on the state space, which would have
resulted in wrong conclusions about the properties of the solutions, since such a compactification would compromise a
treatment of the oscillatory regime at late times.

Although the dynamical system~\eqref{zqsys} is inappropriate for global considerations, we have seen that it
still has advantages, as illustrated by its simplicity and desirable local fixed point properties.
The system has two fixed points located on the $z=0$ boundary, given in~\eqref{FP_qz}.\footnote{For
situating these fixed points in a broader context, see~\cite{barher06}.}
The fixed point $\mathrm{R}$ is a hyperbolic, while $\mathrm{dS}$ is non-hyperbolic with one negative
eigenvalue and one zero eigenvalues. To deal with the zero eigenvalue of $\mathrm{dS}$, we apply center
manifold theory. The negative eigenvalue corresponds to a stable subspace $W^s$ given by the invariant
subset $z=0$, which thereby coincides with the tangential stable subset $E^s$, while the center
manifold $W^c$ has a tangential subspace $E^c$, described by:
\begin{subequations}\label{linearanalysis}
\begin{align}
E^s &= \left\{(z,q)|\,\, z=0\right\},\\
E^c &= \left\{(z,q)|\,\, z - 3(1+q) = 0\right\}.
\end{align}
\end{subequations}
To investigate the center manifold $W^c$ we proceed as in the main text.
We adapt the variables to the tangent space $E^c$ by introducing
\begin{subequations}
\begin{align}
v & = z -3(1+q),
\end{align}
\end{subequations}
which implies that in a neighborhood of $\mathrm{dS}$ the center manifold is described by
the graph $v=h(z)$. From~\eqref{zqsys} it follows that $h(z)$ obeys the first order
differential equation
\begin{equation}
\frac{2}{3}z(z-h(z))\left(\frac{dh}{dz} - 1\right) + 3h(z) + (z-h(z))^2 = 0,
\end{equation}
which is solved approximately for $h$ by a formal series expansion, which
results in
\begin{equation}
q(z) = -1 + \frac{z}{3}\left[1 + \frac{z}{3} - \frac{z^3}{3^5} + \dots\right],
\end{equation}
as $z\rightarrow0$.

Let us now consider an example of another dynamical systems treatment of $f(R)$ flat RW
cosmology which can be found in~\cite{LivRev10}, where the following variables
(when restricted to the present vacuum case) are defined:
\begin{equation}\label{xvardef}
x_1 = -\frac{2\alpha \dot{R}}{H(1 + 2\alpha R)}, \qquad x_2 =
-\frac{R(1 + \alpha R)}{6H^2(1+2\alpha R)}, \qquad x_3 = \frac{R}{6H^2} = 1 - q.
\end{equation}
Viewing this as a variable transformation from $(x_1,x_2,x_3)$ to $(H,\dot{R},R)$ leads to the Jacobian determinant
\begin{equation}\label{xJdet}
\frac{\alpha^2 R^2}{9H^6(1+2\alpha R)^3}.
\end{equation}
The variables thereby break down at $R=0$, $H=0$, and $F=1+2\alpha R=0$, i.e., at
the boundary of the state space of the Einstein frame. It brings parts of the future
null infinity of the state space to finite values of the variables, but other regions,
such as the generic one close to the Minkowski fixed point, are now shifted to infinity
in these unbounded variables.
In the present case it follows that the auxiliary quantity $m$ in~\cite{LivRev10} is given by
$2(x_2+x_3)/x_3$, which when using a time variable $N=\ln a$ leads to the evolution equations
\begin{subequations}\label{xsys}
\begin{align}
\frac{dx_1}{dN} &= -1 - x_3 - 3x_2 + x_1^2 - x_1x_3,\\
\frac{dx_2}{dN} &= \frac{x_1x_3^2}{2(x_2+x_3)} - x_2(2x_3 - 4 - x_1),\\
\frac{dx_3}{dN} &= -\frac{x_1x_3^2}{2(x_2+x_3)} - 2x_3(x_3-2),
\end{align}
\end{subequations}
subjected to the constraint
\begin{equation}
1 = x_1 + x_2 + x_3,
\end{equation}
which can be solved globally for one of the variables. It is clear that not only has 
the system an unbounded incomplete state space, but the equations are also irregular 
as the right hand side blows up on the line $x_2+x_3=0$, $x_1=1$. To conclude, the system is inappropriate
for a global analysis of the problem, but it is still possible to do some local analysis.
To accomplish this, let us solve for one of the variables, e.g., $x_1=1-x_2-x_3$ (which
variable we solve for does not change any conclusions), and look for fixed points. 
Since $x_2+x_3$ appears in the denominator (writing the right hand sides of $dx_2/dN$ and
$dx_3/dN$ with $x_2+x_3$ as a common denominator shows that it is impossible to get rid of
this denominator), the equations are not defined for $x_2+x_3=0$ and hence fixed points
must have $x_2+x_3 \neq 0$.

There are two fixed points: $P_1$: $(x_2,x_3)=(-1,2)$ and $P_4$: $(x_2,x_3)=(5,0)$.
The vacuum fixed point $P_3$ in~\cite{LivRev10} is not defined for the present models,
since it is associated with numerators and denominators in the equations that simultaneously are zero.
Furthermore, $P_6$ in~\cite{LivRev10} only exists when it coincides with $P_1$, for which $m=1$.
Note that the definitions~\eqref{xvardef} implies that $P_1$ corresponds
to the asymptotic limit $H\rightarrow \infty$ and $R\rightarrow \infty$.
%
%
In fact
\begin{equation}
-1=x_2=-x_3\frac{1+\alpha R}{1+2\alpha R}, \qquad 2=x_3=\frac{R}{6H^2},
\end{equation}
which implies $R\rightarrow \infty$ and  $H \rightarrow \infty$. The fixed
point $P_4$ on the other hand implies that $H\rightarrow \infty$ and $R=-1/2\alpha$. This is because
\begin{equation}
5=x_2=-x_3\frac{1+\alpha R}{1+2\alpha R}, \qquad 0=x_3=\frac{R}{6H^2},
\end{equation}
which implies that
$1+ 2\alpha R \rightarrow 0^{-}$ and hence  $H\rightarrow \infty$. Thus both fixed points correspond to part of future null
infinity of the state space light cone. Note that $P_4$ is associated with the boundary of the
Einstein frame.

The fixed point $\mathrm{P}_1$ corresponds to $q=-1$ and is
a non-hyperbolic fixed point with one negative value and one zero eigenvalue with an
associated center manifold, corresponding to the $\mathrm{dS}$ fixed point for $z$ and $q$.
It therefore gives similar results, but the more complicated dynamical system leads to
unnecessary technical complications. The situation for $q=1$ when $H\rightarrow \infty$, is,
however, worse. In this case $\mathrm{P}_4$ is a hyperbolic saddle,
which is associated with a coordinate singularity due to the break down of~\eqref{xJdet}
at $F=1+ 2\alpha R =0$. Because of this breakdown, $\mathrm{P}_4$
yields a solution that comes from a particular direction from future null infinity of the
physical state, namely $R\rightarrow -1/2\alpha$, thereby missing the one-parameter set of
solutions that originates from there into the physical state space. Thus a fixed point
analysis in these variables does not show that there actually is a one-parameter set of
solutions that originate from the limit $q=1$ and $H \rightarrow \infty$, which covers
all solutions except the single solution from $\mathrm{dS}$.

Next we comment on a previous attempt to provide a compact state space, given in~\cite{abdetal12}.
In this work it was assumed that $R>0$ and $F=df/dR>0$, where the latter follows from the first condition
in our case. The variables the authors introduced were given by
\begin{equation}
x = \frac{3\dot{F}}{2fD}, \qquad y = \frac{3f}{2D^2F}, \qquad z= \frac{3R}{2D^2},
\qquad Q = \frac{3H}{D},
\end{equation}
where $D = \left[3\left(H + \frac{\dot{F}}{2f}\right)^2 + \frac{3f}{2F}\right]^{1/2}$.
For the vacuum case one can solve for $y$ and $z$ to obtain a system of evolution equations for
$x$ and $Q$. The authors also introduce an auxiliary quantity $\Gamma$, which for the present
case can be written as
\begin{equation}
\Gamma = \frac{1-x^2}{2Q(Q+2x)}.
\end{equation}
Explicitly inserting this into the equations in~\cite{abdetal12} (which we refrain from giving because of
their considerable complexity) shows that they have $Q(Q+2x)$ in the denominator, which
means that the equations are non-regular and break down at $Q=0$ and at $Q+2x=0$. This result, 
in combination with the fact that the variables only compactify the  $R>0$ part of the state space,
and that $R=0$ is not an invariant subset on the physical state space (except for at the Minkowski fixed point),
unfortunately leads to complications and some erroneous conclusions (due to a breakdown of the time
variable in~\cite{abdetal12}). For example, as we have proven,
all solutions pass through $R=0$ infinitely many times, in contrast to what is claimed in~\cite{abdetal12}.
This example illustrates that compactifications must respect the structure of the state space; if one chooses
to compactify only part of it there will be coordinate singularities associated with the boundary one
chooses for the compactification, unless it is associated with an invariant subset in the original Jordan
state space for $(H,\dot{R},R)$.

\end{appendix}

\end{document}